\pdfoutput=1
\documentclass[12pt,shortbibliography]{article}
\usepackage{authblk}
\usepackage[a4paper,left=2cm,right=2cm,top=2cm,bottom=2cm]{geometry}
\usepackage[english]{babel}
\usepackage[utf8]{inputenc}
\usepackage{amsmath}
\usepackage{amssymb}
\usepackage{wasysym}
\usepackage{graphicx}
\usepackage[colorinlistoftodos]{todonotes}
\usepackage{setspace}
\usepackage{caption}
\usepackage{float}
\usepackage{afterpage}
\usepackage{gensymb}
\usepackage[section]{placeins}
\usepackage{upgreek}
\usepackage{siunitx}
\usepackage{dcolumn}
\usepackage{bm}
\usepackage{epstopdf}
\usepackage{xr}
\usepackage{amsmath}
\usepackage{color}
\usepackage[normalem]{ulem}
\usepackage{wasysym}
\usepackage{siunitx}
\usepackage{physics}
\usepackage{lineno}
\usepackage[T1]{fontenc}
\usepackage{booktabs}
\usepackage[section]{placeins}
\usepackage{textcomp}

\usepackage{gensymb}
\usepackage{xcolor}
\usepackage{multirow}
\usepackage[resetlabels,labeled]{multibib}
\usepackage[explicit]{titlesec}

\usepackage{tabularx}

\usepackage{hyperref}
\hypersetup{colorlinks=true,%
citecolor=blue,%
filecolor=black,%
linkcolor=black,%
urlcolor=blue
}

\UseRawInputEncoding

\hypersetup{
    colorlinks,
    citecolor=blue,
    filecolor=black,
    linkcolor=blue,
    urlcolor=blue
}

\makeatletter

\newcommand{\Rmnum}[1]{\expandafter\@slowromancap\romannumeral #1@}
\makeatother

\title{High-temperature Josephson diode}

\author[1]{Sanat Ghosh\thanks{sanatghosh1996@gmail.com}}

\author[1]{Vilas Patil}

\author[1]{Amit Basu}

\author[1]{Kuldeep}

\author[1]{Achintya Dutta}

\author[1]{Digambar A. Jangade}

\author[1]{Ruta Kulkarni}

\author[1]{A. Thamizhavel}

\author[2]{Jacob F. Steiner}

\author[2]{Felix von Oppen}

\author[1]{Mandar M. Deshmukh\thanks{deshmukh@tifr.res.in}}

\affil[1]{Department of Condensed Matter Physics and Materials Science, Tata Institute of Fundamental Research, Homi Bhabha Road, Mumbai 400005, India}

\affil[2]{Dahlem Center for Complex Quantum Systems and Fachbereich Physik, Freie Universit\"at Berlin, 14195 Berlin, Germany}

\date{}

\begin{document}

\maketitle

\renewcommand{\figurename}{\textbf{Fig.}}



\section*{Abstract}

Symmetry plays a critical role in determining various properties of a material. Semiconducting \textit{p-n} junction diodes exemplify the engineered skew electronic response and are at the heart of contemporary electronic circuits. The non-reciprocal charge transport in a diode arises from doping-induced breaking of inversion symmetry. Breaking of time-reversal, in addition to inversion symmetry in some superconducting systems, leads to an analogous device -- the superconducting diode. Following the pioneering first demonstration of the superconducting diode effect (SDE) \cite{ando_observation_2020}, a plethora of new systems showing similar effects have been reported \cite{zhang_nonreciprocal_2020,wakatsuki_nonreciprocal_2017,baumgartner_supercurrent_2022,lyu_superconducting_2021,bauriedl_supercurrent_2022,narita_field-free_2022,wu_field-free_2022,lin_zero-field_2022,trahms_diode_2023}. SDE lays the foundation for realizing ultra-low dissipative circuits, while Josephson phenomena-based diode effect (JDE) can enable realization of protected qubits \cite{larsen_parity-protected_2020,schrade_protected_2022}. However, SDE and JDE reported thus far are at low temperatures ($\sim 4$~K or lower) and impede their adaptation to technological applications. Here we demonstrate a Josephson diode working up to 77~K using an artificial Josephson junction (AJJ) of twisted layers of Bi$_2$Sr$_2$CaCu$_2$O$_{8+\delta}$ (BSCCO). The non-reciprocal response manifests as an asymmetry in the magnitude of switching currents as well as their distributions, and appears for all twist angles.
The asymmetry is induced by and tunable with a very small magnetic field applied perpendicular to the junction. We report a record asymmetry of 60~\% at 20~K. We explain our results within a vortex-based scenario. Our results provide a path toward realizing superconducting quantum circuits at liquid nitrogen temperature.


\section*{Introduction}

Breaking of inversion and time-reversal symmetry is typically required to realize SDE \cite{tokura_nonreciprocal_2018}. 
Several proposals for realizing the superconducting analogue of the diode effect \cite{hu_proposed_2007,chen_asymmetric_2018,misaki_theory_2021} have been explored. It was followed by successful demonstrations in various systems such as non-centrosymmetric superconductors \cite{ando_observation_2020,zhang_nonreciprocal_2020,wakatsuki_nonreciprocal_2017}, two-dimensional electron gases \cite{baumgartner_supercurrent_2022}, patterned superconductors \cite{lyu_superconducting_2021,bauriedl_supercurrent_2022}, superconductor/ferromagnet multilayers \cite{narita_field-free_2022,wu_field-free_2022}, and twisted graphene systems \cite{lin_zero-field_2022,diez-merida_magnetic_2021}. Although several different mechanisms \cite{wakatsuki_nonreciprocal_2017,yuan_supercurrent_2022,daido_intrinsic_2022,davydova_universal_2022,zhang_general_2022,tanaka_theory_2022} have been proposed to describe the observations across systems, the theories are still at their early stage of development. The SDE/JDE, which is the result of non-reciprocal response in these systems, manifests in terms of different magnitudes of superconducting switching currents ($I_\text{s}^{+} \neq I_\text{s}^{-}$) in the two opposite polarities. The effect has been demonstrated as magnetic field induced \cite{ando_observation_2020,lyu_superconducting_2021,bauriedl_supercurrent_2022,wakatsuki_nonreciprocal_2017,zhang_nonreciprocal_2020,baumgartner_supercurrent_2022} and field free \cite{narita_field-free_2022,wu_field-free_2022,lin_zero-field_2022} in nature. In terms of practical usability, however, all reports to date require very low temperatures, 4~K or less, to operate. We overcome this challenge using \textit{c-axis} Josephson junctions (JJ) between flakes of a high-$T_c$ cuprate superconductor.


We demonstrate JDE in an artificially created Josephson junction (AJJ) by twisting two layers of BSCCO \cite{maeda_new_1988} that can be exfoliated to atomically flat flakes \cite{liao_superconductorinsulator_2018,sterpetti_comprehensive_2017,zhao_sign-reversing_2019,ghosh_-demand_2020} sustaining superconductivity down to the monolayer limit \cite{yu_high-temperature_2019}. The difference between the magnitude of positive ($I_\text{s}^{+}$) and negative ($I_\text{s}^{-}$) switching currents and their statistical distributions, in these junctions, are tunable with a very small magnetic field applied perpendicular to the plane of the junction. The asymmetry persists from low temperature up to the superconducting transition temperature ($T_\text{c}$, $\sim$ 77 K) of the junction and increases as temperature decreases, reaching a value as high as $60~\%$ at 20~K. The diode behavior is demonstrated in terms of the half-wave rectification of a square wave current.


Heterostructures, assembled by twisting two layers of van der Waals materials offer a new platform for emergent electronic responses, entirely different from the constituent materials. In a similar spirit, a recent study \cite{can_high-temperature_2021} proposes the possibility of realizing time reversal symmetry broken high-temperature topological superconductivity at the interface of two 45\degree~twisted BSCCO layers. Thus twisted BSCCO JJs offer a natural platform to explore the physics of JDE. Notably, our work demonstrates that JDE is a feature of the artificial \textit{c-axis} JJs and does not require a special angle, 45\degree, as long as inversion symmetry is broken. 


Fabricating twisted BSCCO junctions, however, is extremely challenging, and many of the junction properties are sensitive to fabrication methods. In the past, there have been numerous studies on artificially created twisted junctions of BSCCO to study the pairing symmetry of the superconducting order parameter. But the fabrication process involved high-temperature oxygen annealing to sustain the superconductivity at the interface. This led to observations of no \cite{li_mathrmbi_2mathrmsr_2mathrmcacu_2o_8ensuremathdelta_1999,latyshev_c-axis_2004,zhu_presence_2021} or different \cite{takano_d-like_2002,lee_twisted_2021} angular dependence of the Josephson coupling, different than the anticipated $d$-wave superconductivity in BSCCO \cite{shen_anomalously_1993,gu_directly_2019}. Recently, there has been progress in making twisted BSCCO JJs using room temperature exfoliation \cite{zhu_presence_2021,lee_twisted_2021} and employing cryogenic exfoliation \cite{zhao_emergent_2021}. 

We fabricate twisted BSCCO JJs following Zhao \textit{et al.} \cite{zhao_emergent_2021}. The junctions are created by re-exfoliating a relatively thicker BSCCO flake into two pieces and stacking them, which ensures alignment of the crystal axis and a well-defined twist angle (with an accuracy of $\pm~0.5\degree$). The re-exfoliation is done inside an Ar-filled glove box with a low-temperature stage. To avoid any chemical contamination, Au contacts are directly deposited on the flake through a pre-aligned SiN mask. More details of the device fabrication are discussed in the methods section.


\section*{Experimental data}

\begin{figure*}[h]
\centering

\includegraphics[width=17cm]{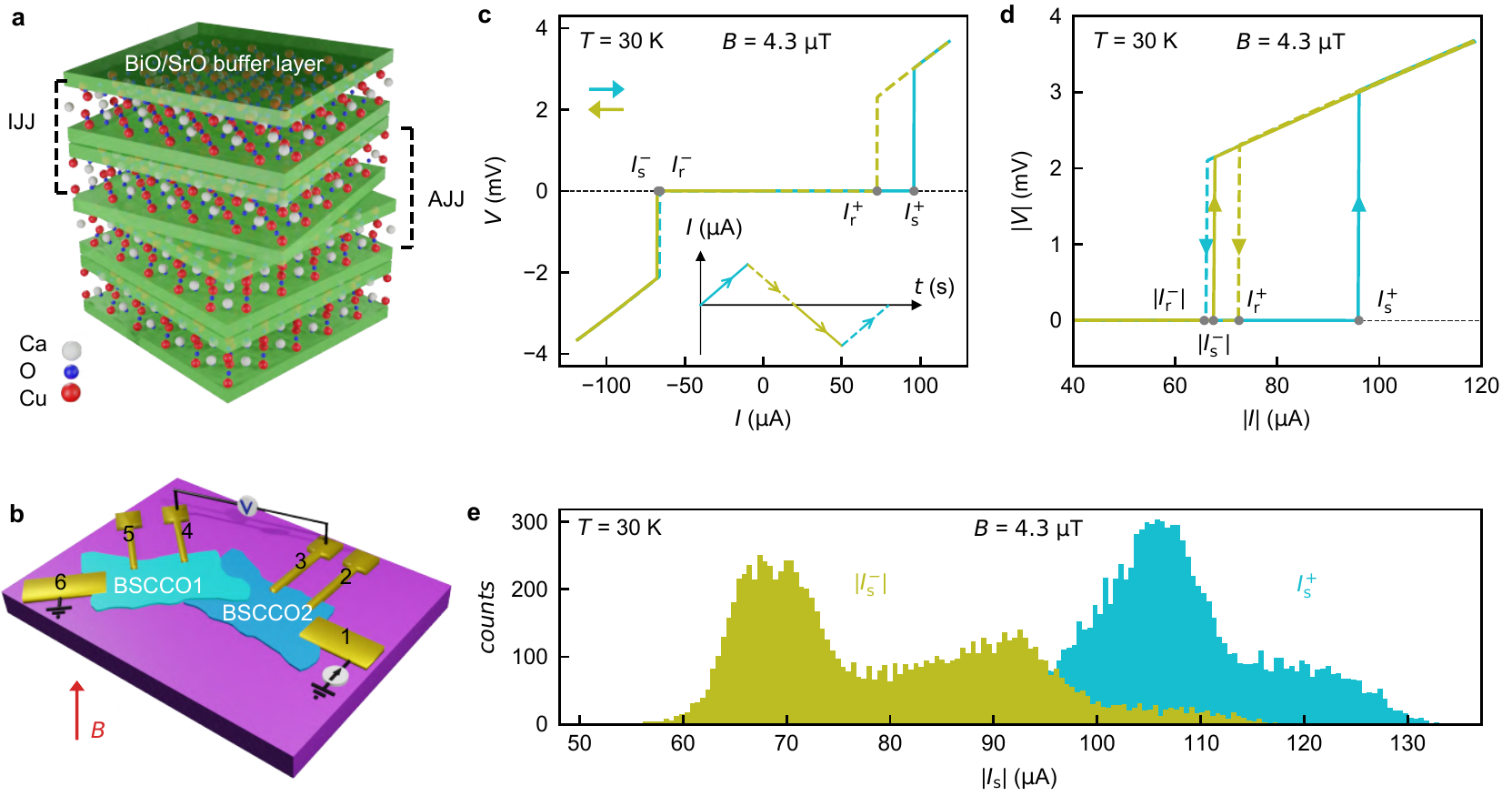}
\caption{\label{fig:fig1}\textbf{Asymmetry of switching currents in a 45\degree~twisted artificial Josephson junction of BSCCO.} (a) Schematic of twisted BSCCO (not to scale). One unit cell of BSCCO is rotated at a specific angle with respect to the other unit cell. In each unit cell, the Cu-O planes are separated by insulating SrO/BiO buffer layers (green slabs) and constitute an intrinsic Josephson junction (IJJ). At the interface of the twisted region, an artificial Josephson junction (AJJ) is formed. (b) Schematic of the twisted BSCCO device with Au contacts for measurements. Electrodes 1 and 6 are used as source and sink for the current, and properties across the twisted junction are measured with electrodes 3 and 4. Magnetic field $B$ is perpendicular to the junction plane. (c) \textit{dc} $I-V$ characteristic across the junction at 30~K. The two different colored line plots are for two different sweep directions of the biasing current, indicated by the arrows. Switching currents ($I_\text{s}$) and retrapping currents ($I_\text{r}$) for positive and negative bias regimes are marked with a superscript. The current through the device is swept in the order as shown in the inset. (d) Absolute values of $I$ and $V$ for both positive and negative bias range for the same data in (c) at 30~K. The solid cyan and green curves are for switching branch. The dashed cyan and green curves are for the retrapping branch. This plot clearly shows the asymmetry in the magnitude of $I_\text{s}^{+}$ and $I_\text{s}^{-}$. (e) Histograms of the switching currents, $I_\text{s}^{+}$ and $I_\text{s}^{-}$ at 30~K. $10^4$ switching events were taken to get the distributions. The triggering voltage used for measuring switching currents was $\pm 0.2$ mV. The median values of these distributions also show asymmetry between the magnitude of $I_\text{s}^{+}$ and $I_\text{s}^{-}$. }

\end{figure*}

Fig.~\ref{fig:fig1}a and b show the schematic of the twisted BSCCO crystal structure and the device geometry of the artificial JJ, respectively. The crystal structure of BSCCO is such that it has its own IJJ. The superconducting Cu-O planes in this material, separated by the insulating SrO/BiO buffer layers (green slabs in Fig.~\ref{fig:fig1}a), are Josephson coupled and constitute a series of IJJs in the material structure \cite{kleiner_intrinsic_1992}. By twisting two individual layers of BSCCO, we create an AJJ at the interface, as indicated in Fig.~\ref{fig:fig1}a. The twist angle controls \cite{lee_twisted_2021,zhao_emergent_2021} the switching current density ($J_\text{s}$), and consequently the Josephson energy, of the AJJs (maximum for 0\degree, and minimum for 45\degree~twist) due to $d$-wave symmetry of the superconducting order parameter in the system.

We made 10 AJJs with different twist angles (four 45\degree, four 0\degree, and two 22\degree). In the main text, we present data of one of 45\degree~(D1) and 0\degree~(D2) twisted junctions, and data of other twisted junctions are shown in Supplementary Information Fig. S6 and Extended Data Fig. 4 (22\degree, D5). The JDE is observed in all of the devices with different twist angles, an important aspect of our work. We design our devices such that we can simultaneously measure the junction properties and the two flakes that make the junction (Fig.~\ref{fig:fig1}b). We do this to ensure that there is no degradation during the fabrication. First, we check the four terminal resistance across the junction as a function of  temperature  which shows $T_\text{c}$ similar to the pristine BSCCO flake, as shown in the Supplementary Information, Fig. S3. The pristine nature of the fabricated junctions is evident from the fact that the 0\degree~twisted junction has magnitude of switching current density ($J_\text{s}$) similar to the IJJs formed between the Cu-O planes in BSCCO (see Extended data Fig. 1).  


Fig.~\ref{fig:fig1}c shows $dc$ $I-V$ across the junction at 30~K and at fixed $B = 4.3$ \textmu T, perpendicular to the junction plane (see Fig.~\ref{fig:fig1}b). Temperature dependence of $dc$ $I-V$ is shown in Supplementary Information Fig. S11. We discuss the detailed field dependence later. The two colored line plots are for different sweep directions of the bias current as indicated by the arrows. The bias current is swept in the order as shown in the inset. We identify switching ($I_\text{s}$) and retrapping currents ($I_\text{r}$) in positive and negative bias range with $I_\text{s}^{+}$ and $I_\text{s}^{-}$, as shown in Fig.~\ref{fig:fig1}c. The magnitude of $I_\text{s}^{+}$ is different from $I_\text{s}^{-}$. The observed hysteresis between $I_\text{s}$ and $I_\text{r}$ is a well-understood physics of an underdamped JJ \cite{tinkham_introduction_2004}.

To clearly see the difference between the two switching currents $I_\text{s}^{+}$ and $I_\text{s}^{-}$, we plot the $I-V$ curves of the negative bias branch by flipping their signs in Fig.~\ref{fig:fig1}d. From the plot, it is clear that $I_\text{s}^{+}$ is larger than $\abs{I_\text{s}^{-}}$. This means that at bias currents $I$, $\abs{I_\text{s}^{-}} < I < I_\text{s}^{+}$, the system will dissipate energy and act like a resistive element in negative bias and be superconducting in positive bias. To quantify the difference between $I_\text{s}^{+}$ and $I_\text{s}^{-}$ we define an asymmetry factor, $\alpha$ as $\alpha = (I_\text{s}^{+} - \abs{I_\text{s}^{-}})/(I_\text{s}^{+} + \abs{I_\text{s}^{-}}) \times~100~\%$. The value of $\alpha$ for the data, shown in Fig.~\ref{fig:fig1}d at 30~K, is $17~\%$.

Switching of JJs from superconducting to resistive state is stochastic and has a finite distribution at a fixed temperature \cite{fulton_lifetime_1974}. To probe the switching statistics for $I_\text{s}^{+}$ and $I_\text{s}^{-}$ in our twisted JJ, we measure $10^4$ switching events for both positive and negative bias currents and compare their distributions. The measurement protocol for switching distribution is discussed in the methods section. As seen in Fig.~\ref{fig:fig1}e, the median value of the distribution also shows asymmetry, as was seen in a single sweep. In addition, as we tune $B$, the spread of the distributions becomes asymmetric -- an aspect we discuss later. 


\begin{figure*}[h]
\centering

\includegraphics[width=17cm]{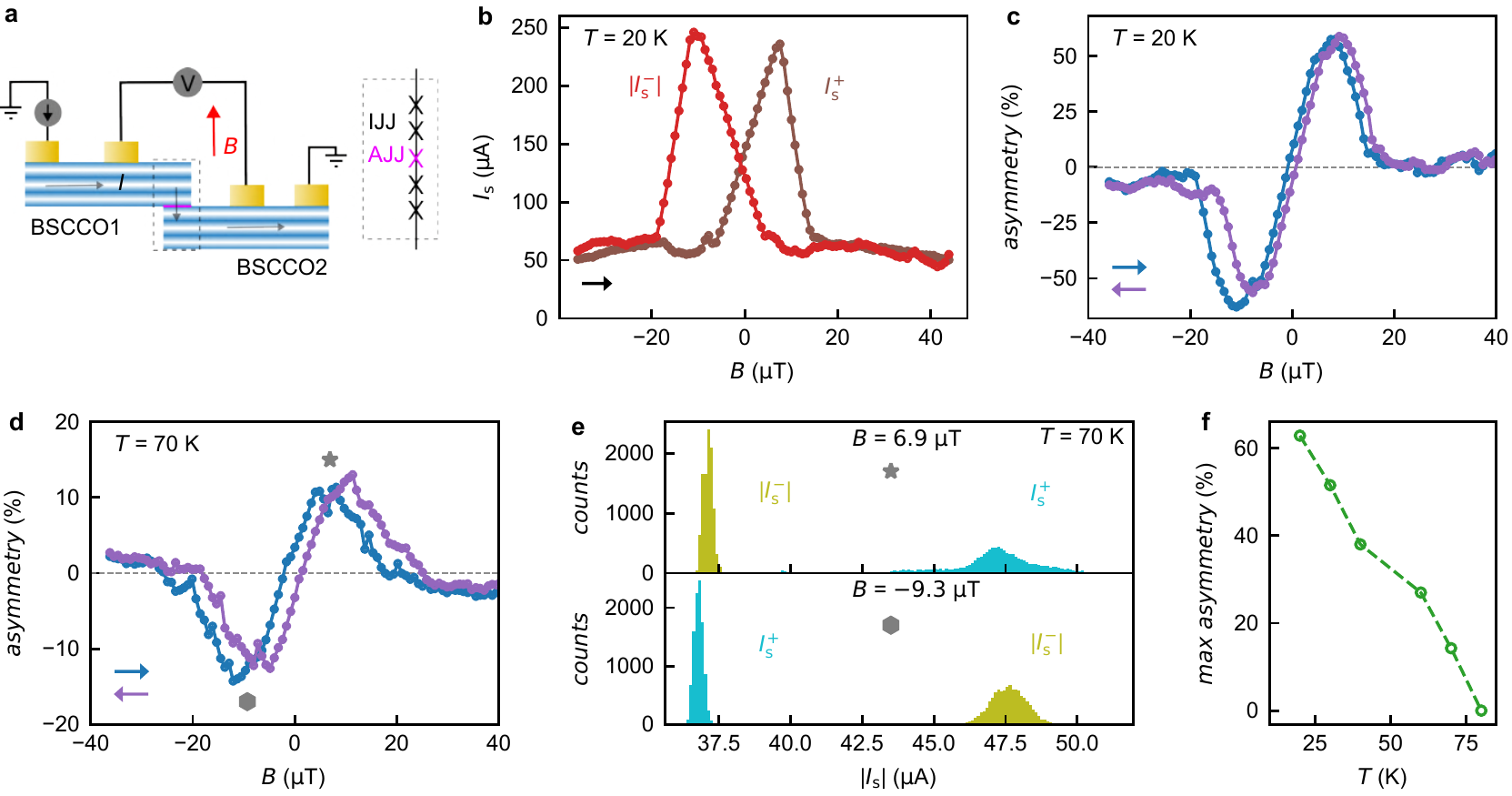}
\caption{ \label{fig:fig2}  \textbf{Switching current asymmetry with perpendicular magnetic field and temperature for the 45\degree~twisted BSCCO junction.} (a) Cross-sectional schematic of the twisted BSCCO device. At the junction, current ($I$) flows along the c-axis of BSCCO. Applied $B$ is parallel to $I$ but perpendicular to the junction plane. An equivalent picture at the junction comprising of IJJs and AJJ is shown on the right. (b) Variation of switching currents, $I_\text{s}^{+}$ and $I_\text{s}^{-}$ with $B$ at 20~K. For each value of $B$, 100 switching events were recorded with the help of a counter and were averaged to get $I_\text{s}^{+}$ and $I_\text{s}^{-}$. The measurement is done in cryogenic setup 1 with a homemade electromagnet. We have subtracted $-5$ \textmu T from the data due to an offset value of $B$ in the measurement setup (see Supplementary Information, section S13 for details). (c) Asymmetry of switching currents $\alpha$ with the externally applied magnetic field $B$ at 20~K. $\alpha$ is calculated from (b), as described in the main text. The two different colored line plots are for different sweep directions of $B$, indicated by the arrows. (d) Variation of calculated $\alpha$ with $B$ at 70~K from the same device. (e) Distributions of switching currents ($I_\text{s}^{+}$ and $I_\text{s}^{-}$) at 70~K for two different $B$ values, as indicated in (d) by star and hexagon. The spread of the distributions changes by the change in sign of $B$. (f) The maximum value of the asymmetry factor as a function of temperature. At each temperature, maximum asymmetry is obtained from the $B$ sweep. The asymmetry persists below $T_\text{c}$ of the junction and increases as the temperature is lowered reaching a value as high as $60~\%$ at 20~K. } 

\end{figure*}


We next study the current asymmetry $\alpha$ at different temperatures and with an external magnetic field $B$, applied perpendicular to the plane of the junction. Fig.~\ref{fig:fig2}a shows the cross-sectional schematic of the device showing the IJJs formed between Cu-O planes and the AJJ, formed artificially. As depicted in Fig.~\ref{fig:fig2}a, the direction of the current across the AJJ is along the c-axis, parallel to applied $B$. Fig.~\ref{fig:fig2}b shows the variation of $I_\text{s}^{+}$ and $I_\text{s}^{-}$ with $B$ at 20~K. At each $B$ value, we record 100 switching events with the help of a counter and average it out to get $I_\text{s}^{+}$ and $I_\text{s}^{-}$, which are plotted in Fig.~\ref{fig:fig2}b. The measurement protocol is discussed in the methods section. Importantly, we note a pronounced change of $I_\text{s}^{+}$ and $I_\text{s}^{-}$ with a very small magnetic field. Variations of $I_\text{s}^{+}$ and $I_\text{s}^{-}$ show peaks that are shifted from each other along the $B$ axis. The difference in $B$ between the two peaks corresponds to a few flux quanta through the junction area. We calculate $\alpha$ from this data. Fig.~\ref{fig:fig2}c,d show the evolution of calculated $\alpha$ with $B$ at 20 and 70~K, respectively. The line plots with different colors are for two sweep directions of $B$ as indicated by the inset arrows. Here, we note the main observation. $\alpha$ is antisymmetric in $B$, $\alpha (B) = - \alpha (-B)$ and peaks at specific fields in both positive and negative directions of $B$. This implies that $I_\text{s}^{+}$ is larger than $\abs{I_\text{s}^{-}}$ at $+B$ and smaller at $-B$. The field modulation of the asymmetry means that the effect we observe is field induced in nature and can be tuned/controlled by a very small magnetic field $\sim$ 10 \textmu T. We have measured this device in cryogenic setup 1 using a homemade electromagnet to generate very small fields (see Supplementary Information, section S12 for details).

Additionally, we find the antisymmetric behavior with $B$ is not restricted to the magnitude of $\alpha$ but is also reflected in the switching distribution widths of $I_\text{s}^{+}$ and $I_\text{s}^{-}$. To see this, we record $10^4$ switching events for specific $B$ values (indicated in Fig.~\ref{fig:fig2}d by star and hexagon). Fig.~\ref{fig:fig2}e shows the distributions of the two switching currents in $+B$ and $-B$ at 70~K. From the plots, we clearly see the spread of the distribution for $I_\text{s}^{+}$ and $I_\text{s}^{-}$ is flipped with flipping the sign of $B$. This behavior is present at all temperatures below $T_\text{c}$ (see Supplementary Information, Fig. S4).

To see how $\alpha$ varies with temperature, we determine its maximum value from the $B$ sweep at each temperature. The extracted value of maximum asymmetry is plotted with temperature in Fig.~\ref{fig:fig2}f. Importantly, the asymmetry persists till very close to $T_\text{c}$ and increases as the temperature is lowered below $T_\text{c}$, reaching a value as high as $60~\%$ at 20~K. This value of asymmetry factor is the highest reported so far at this temperature across systems showing SDE/JDE (see Supplementary Information, Section S4 for a comparison of the asymmetry factors and operating temperatures for different systems).


\section*{Half-wave rectification of the superconducting diode}

To show actual diode-like behavior, we demonstrate the half-wave rectification of an $ac$ current. For that, we send a square wave excitation current of specific magnitude and frequency 0.1~Hz, and the voltage drop across the junction is measured. The magnitude of the current is such that it is larger than one of the switching currents ($I_\text{s}^{+}$ or $\abs{I_\text{s}^{-}}$) and smaller than the other. The results at 50~K are shown in Fig.~\ref{fig:fig3}. Rectification data at other temperatures are shown in Supplementary Information, Fig. S5. We see half-wave rectification of the negative half-cycle of the supercurrent at $B = 4.2$ \textmu T in Fig.~\ref{fig:fig3}a. By an externally applied $B$, the asymmetry of switching currents can be reversed, which in turn changes the rectification nature of the junction. Fig.~\ref{fig:fig3}b shows rectification of the positive half-cycle of the supercurrent in presence of $B =- 7.9$ \textmu T. The rectification ratio, as defined by $V(+I)/V(-I)$, is very high ($\sim5000$) at 50~K.

\begin{figure*}[h]
\centering

\includegraphics[width=9cm]{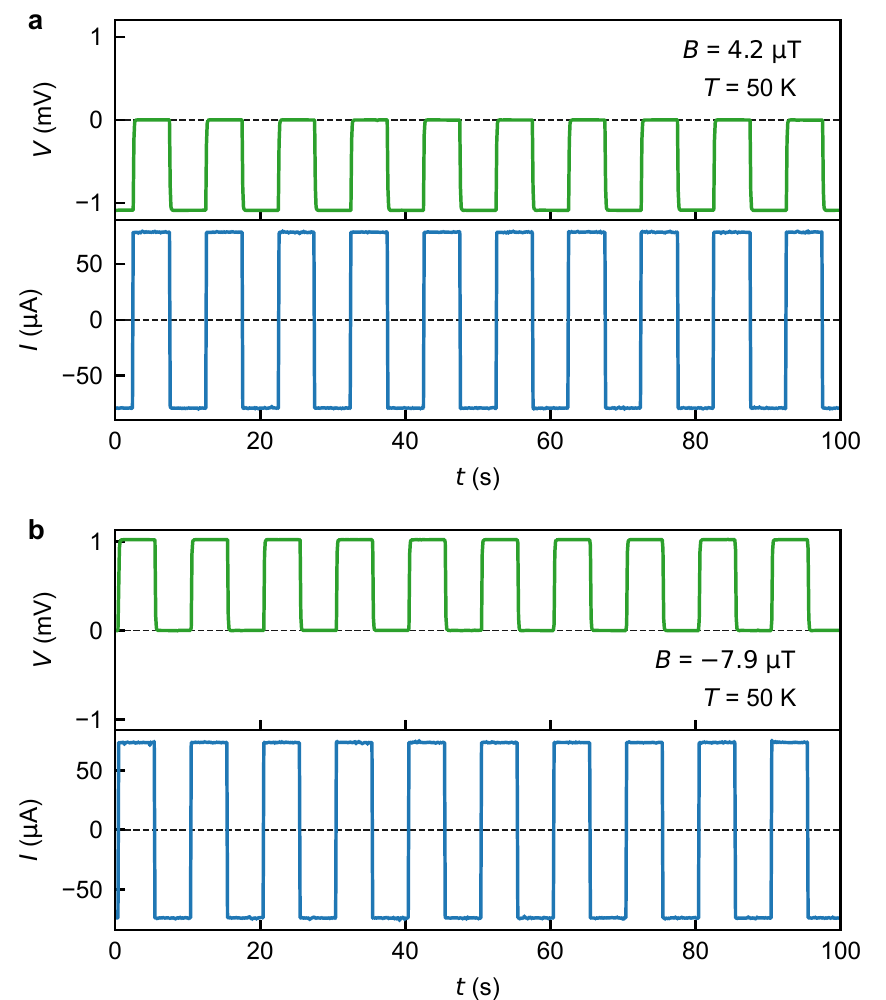}
\caption{ \label{fig:fig3}  \textbf{Half wave rectification of the superconducting diode.} (a) Half wave rectification of the current by the device at $4.2$~\textmu T magnetic field and at 50~K. A square wave excitation current of frequency 0.1~Hz was sent through the device and the voltage drop across the junction was measured. The magnitude of the square excitation current is larger than $\abs{I_\text{s}^{-}}$ and smaller than $I_\text{s}^{+}$ and hence there is a voltage drop only in the negative half-cycle of the current resulting in rectification. (b) Rectification by the device at 50~K but at $-7.9$~\textmu T field. At this field value, $\abs{I_\text{s}^{-}}$ becomes larger than $I_\text{s}^{+}$ and voltage drop across the junction appears in the positive half-cycle of the current. }

\end{figure*}

\section*{Control experiments}


\begin{figure*}
\centering

\includegraphics[width=17cm]{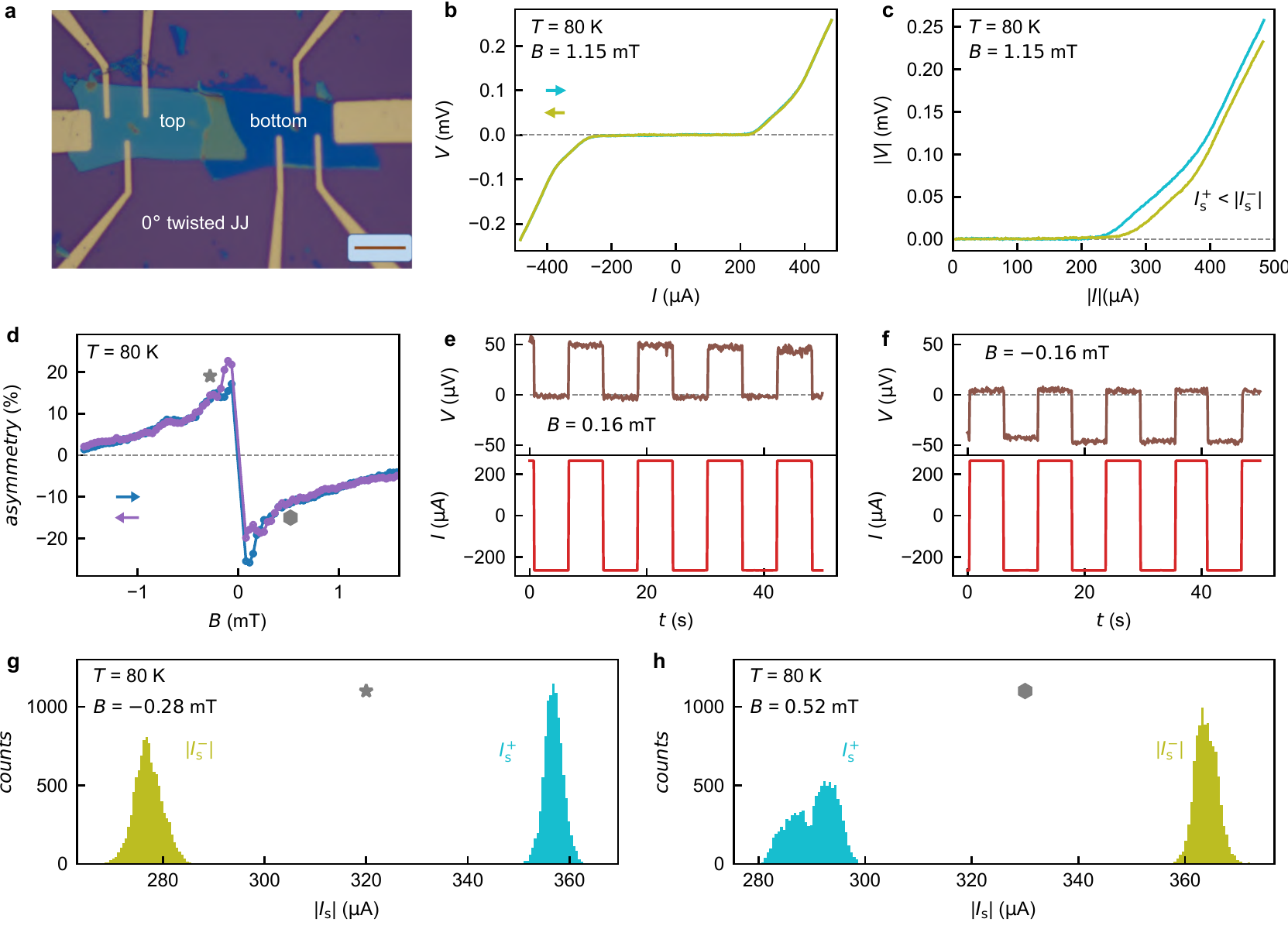}
\caption{ \label{fig:fig4}  \textbf{Superconducting diode behavior of 0\degree~twisted BSCCO Josephson junction.} (a) Optical micrograph image of 0\degree~twisted BSCCO JJ. Scale bar is $20$ \textmu m. (b) $dc$ $I-V$ characteristic of the junction at 80~K. Two curves for up and down sweep of bias current, indicated by the two arrows, sit on top of each other indicating no hysteresis. (c) Absolute values of $I$ and $V$ are plotted to clearly see the asymmetry. Here |$I_\text{s}^{-}$| is larger than $I_\text{s}^{+}$. (d) Variation of asymmetry factor $\alpha$ (defined in the text) with $B$ at 80~K. $B$ is applied perpendicular to the junction plane. For each $B$ value, 100 switching events (trigger voltage $\pm 0.03$ mV) were recorded and averaged to get $I_\text{s}^{+}$ and $I_\text{s}^{-}$. The $\alpha$ is calculated from the average values of $I_\text{s}^{+}$ and $I_\text{s}^{-}$.  The different colored curves indicate two sweeping directions of $B$, as indicated by the arrows. Similar to the 45\degree~device, we see $\alpha$ to be antisymmetric with $B$. The measurement has been done in setup 2 with a superconducting magnet. We have subtracted an offset value of $-1.15$ mT to position the zero of $B$ in between the two extrema of $\alpha$. The offset is likely due to the trapped flux in the superconducting magnet (see Supplementary Information, section S13 for details). (e),(f) Half-wave rectification response of the junction at 80~K for two different $B$ values. Rectification ratio for this device at 80 K is $\sim$ 25. (g),(h) Distributions of the switching currents ($I_\text{s}^{+}$ and $I_\text{s}^{-}$) at two different $B$ values. $10^4$ switching events were recorded to get the histograms.} 

\end{figure*}


All the observations described so far are exclusive to the AJJ at the interface of the two BSCCO flakes. We do several control checks to conclude this. We see negligibly small $\alpha$ ($\sim 0.4~\%$) and no significant effect of $B$ on switching current distributions for the pristine BSCCO flake in the same device away from the junction, as shown in the Extended data Fig. 2. Our device geometry and contacts configuration for the  measurements are such that the measured response across the junction region includes junction properties at the interface as well as contributions from IJJs of bulk BSCCO flake, as seen in Fig.~\ref{fig:fig2}a. However, at smaller bias currents, only the artificial junction contributes by switching from superconducting to a normal state. As the current is increased further, the IJJs come into picture and switch one by one. Crucially, we observe asymmetry in $I_\text{s}$ of AJJs and small asymmetry ($\sim 0.8\%$) of IJJs in the same device. Also, the switching current distributions of IJJ do not show any tunability with applied $B$. The data is shown in the Extended data Fig. 3. From all this evidence, it is very clear that the diode behavior arises from the artificial junction created by twisting two flakes of BSCCO at the interface.

So far, we have discussed the results of a 45\degree~twisted device, but the effect we see is not restricted to this specific twist angle. We observe the diode behavior in a 0\degree~twisted device as well. Fig.~\ref{fig:fig4}a shows the optical image of the 0\degree~twisted device, and Fig.~\ref{fig:fig4}b and c show the asymmetric nature of $I_\text{s}^{+}$ and $I_\text{s}^{-}$. The asymmetry $\alpha$ is tunable with $B$ (Fig.~\ref{fig:fig4}d), as was the case with the 45\degree~twisted device discussed earlier. Fig.~\ref{fig:fig4}e and f show the rectification data at 80~K and its tunability with $B$. In addition, the switching current distribution is tuned by $B$. This indicates the key role of the magnetic field in the underlying mechanism of the JDE. Response of the 0\degree~twisted device implies that the JDE effect is generic to AJJ  at the interface and that the order-parameter symmetry does not play a significant role. However, the order parameter controls the magnitude of the switching current for different twist angles. The data from the 0\degree~twisted device, where our alignment accuracy is $\pm~0.5\degree$, suggests that even the inadvertent breaking of interfacial inversion symmetry is crucial. As an additional control test, we fabricated a natural step-edge device where we see small switching current asymmetry (see Supplementary Information, Fig. S7 and S8). Moreover, the asymmetry is largest in $45\degree$ twisted device and decreases with a decrease in twist angle (see Extended Data Fig. 5).

Importantly, we observe large asymmetry in our devices for out-of-plane $B$ and no asymmetry for in-plane $B$ (see Supplementary Information Fig. S13). For large in-plane $B$ we observe a typical Fraunhofer pattern for a JJ (see Supplementary Fig. S12). We note the out-of-plane $B$ range over which we see asymmetry tunability is smaller in $45\degree$ twisted devices than $0\degree$ devices (see table in section S11 in Supplementary Information).


\section*{Mechanisms for SDE}

We now discuss possible mechanisms underlying the JDE in twisted BSCCO JJ (see Extended data Fig. 6 for a schematic and 
Supplementary Information section S17 for additional details). The data indicate  that the diode response originates from the interface between top and bottom crystals (AJJ). This interface breaks inversion symmetry due to misalignment of crystal axes as well as uncorrelated defect distributions. Moreover, the diode response occurs when  applying a magnetic field, which breaks time-reversal symmetry. 
Along with the observation of a stronger asymmetry in switching than  retrapping, this suggests an origin in the current-phase relation \cite{steiner_diode_2022}. 

We estimate that the Josephson penetration depth ($\sim 100$ \textmu m) is large compared to the width of the junctions \cite{zhao_emergent_2021}. In this situation, the spatial dependence of the gauge-invariant phase difference across the AJJ is governed by the externally applied magnetic field. The phase difference depends on the magnetic-field component parallel to the $ab$ plane as well as differences in vortex configuration between the top and bottom flakes. The critical current of the junction is governed by the resulting Fraunhofer-like interference \cite{tinkham_introduction_2004}.

In the experiment, the field is applied along the $c$-axis. As the London penetration depth $\lambda_{ab}\sim 0.2$ \textmu m is small compared to the width of the flakes, this  field induces stacks of pancake vortices in the top and/or bottom flakes of the junction. The stacks will tend to pin at different locations in top and bottom flake due to uncorrelated defect configurations. At the AJJ, the magnetic flux of these vortex stacks either escapes or connects (via a Josephson vortex) within the $ab$ plane \cite{golubov_interaction_1992, blatter_vortices_1994,grigorenko_one-dimensional_2001,cole_ratchet_2006}. The difference in the vortex configurations as well as the field along the $ab$-plane (Josephson vortex) generate a spatially dependent phase difference, affecting the critical current of the junction (see Extended data Fig. 6).  
From the sample geometry, we see the bias current has a component flowing in the $ab$ plane. This current induces an asymmetry in the switching current due to the associated magnetic field in the $ab$ plane and the resulting Magnus force acting on the stacks of pancake vortices. The induced field modifies the Fraunhofer interference asymmetrically in bias-current direction \cite{golod_demonstration_2022}. We estimate a resulting asymmetry in the percent range, considerably below the maximal observed asymmetries (see Supplementary Information section S18C for details). Larger asymmetries can be induced by Magnus forces on the vortex stacks by the bias current. This force acts in opposite directions for the two bias directions, inducing significant and asymmetric rearrangements of the vortex configuration, with a concurrent change of the Fraunhofer interference. 

This scenario is consistent with the central observations: (i) The asymmetry sets in at low fields when the first vortices enter along the $c$-axis. With increasing $B$, the variation in phase difference increases along with the number of vortices, reducing the amplitude of the Fraunhofer interference and hence the asymmetry. (ii) Fields along the $c$-axis are effectively focused into the $ab$-plane. No corresponding focusing occurs for fields applied along the $ab$-plane due to the much larger $\lambda_c$ and smaller sample cross-section. This explains the absence of asymmetry when applying the field along the $ab$-plane. 
(iii) With increasing twist angle, the Josephson coupling between vortex stacks in top and bottom flakes weakens due to the $d$-wave nature of superconductivity. This implies weaker correlations between vortices in the bottom and top flakes and thus larger variations in the phase difference, explaining the trend to lower onset fields as the twist angle increases.
(iv) Vortex pinning induces a weakly hysteretic $B$-field dependence of the asymmetry, as regularly observed. (v) Multistabilities of the vortex configuration induce asymmetry and multiple humps in the distribution of the switching current. 

Vortex-induced mechanisms have also been discussed for diode effects in the bulk critical current of thin-film superconductors \cite{hou_ubiquitous_2022,brandt_thin_2005}. 
While the critical current of our junctions is governed by Fraunhofer interference, it is controlled by vortex flow (and asymmetric edge barriers for vortices \cite{vodolazov_superconducting_2005,sivakov_spatial_2018,suri_non-reciprocity_2022}) in the case of the superconducting thin films. We note that a possible edge barrier effect is very small in our samples as we observe only weak asymmetry ($\sim 0.4$ \%) in pristine BSCCO flakes in the same device away from the junction. We can also eliminate a contribution coming from magnetochiral anisotropy (MCA) \cite{ando_observation_2020} in our case as we see JDE with $B$ parallel to both the current flowing directions ($J$) and the axis of  broken inversion symmetry at the junction.


\section*{Conclusions}

In summary, we report and explain the Josephson diode effect in artificially created Josephson junctions of twisted  BSCCO. Our work demonstrates the highest asymmetry in magnitudes of switching currents reported so far of $60~\%$ at 20~K. While the magnitude of the switching current is tuned by twist angle and sets the Josephson energy of these junctions, the asymmetry on the other hand is tuned by an independent knob -- a very small magnetic field ($\sim$ 10 \textmu T) applied perpendicular to the junction plane. The asymmetry persists even at 80~K for some twist angles. Demonstration of Josephson diode effect above liquid nitrogen boiling point (77~K) is a significant advancement that will ease its adaptation to real applications in circuits with ultra-low dissipation. The required small fields to tune asymmetry can be generated on-chip allowing greater design flexibility. Protected qubits \cite{larsen_parity-protected_2020,schrade_protected_2022} based on the Josephson diode operating at higher temperatures could lead to a new architecture for quantum devices.

\clearpage


\section*{Methods}

\subsection*{Device fabrication}

We developed the cryogenic exfoliation setup for fabrication of the BSCCO Josephson junctions. It is a completely dry pick-up and transfer method and can be employed to exfoliate various 2D materials. The cryogenic exfoliation technique allows re-exfoliation of a relatively thicker BSCCO flake, which is pre-exfoliated on a Si/SiO$_2$ chip. Re-exfoliation from the same flake ensures the alignment of the crystal axis. The re-exfoliated flake is rotated by a specific angle and immediately transferred onto the same flake to form a twisted junction. The low-temperature exfoliation method is very efficient in preventing the degradation of interfacial superconductivity.

Inside an Ar-filled glove box, we exfoliate BSCCO from an optimally doped bulk crystal by scotch tape technique on plasma-cleaned Si/SiO$_2$ substrate. The Si/SiO$_2$ substrate is heated inside the Ar-filled glove box overnight before exfoliation to remove any adsorbed gases. The substrate with exfoliated BSCCO is then placed on a cold stage under a microscope. We pass liquid nitrogen through the stage to cool it. Once the temperature of the stage reaches $-25\degree$~C, we attach hemispherical polydimethylsiloxane (PDMS) stamp with the appropriately identified thicker BSCCO flake. The temperature of the stage is lowered further. At around $-90\degree$~C the PDMS stamp goes through its glass transition point. Consequently, it gets detached automatically from the substrate by partially cleaving thick BSCCO flake into two pieces. Next, we quickly rotate the upper stage on which PDMS is mounted by a specified angle (with $\pm~0.5\degree$ accuracy) and re-assemble it. The cold stage is then warmed slowly to $10\degree$~C and the hemispherical stamp is detached. After the stack reaches room temperature, we take it out and align it with a SiN mask inside a clean room for contact deposition. 70~nm Au contacts are deposited by e-beam evaporation. For more details on cryogenic exfoliation, see Supplementary Information, Section S1. The thicknesses of all measured devices are tabulated in Supporting Information, section S11.

\subsection*{Measurements}

\subsubsection*{\textit{dc} \textit{I-V} measurements}

$dc$ voltage from NI DAQ is fed to a series resistor (10~k$\Omega$). The outcoming $dc$ current is sent through the device by the current injecting electrode. The current is then collected by a current-to-voltage converter and measured with NI DAQ. The voltage drop across the device is also measured with the DAQ after amplification by a voltage preamplifier.  

\subsubsection*{Measurement of switching current distributions}

For distributions of switching currents with $10^4$ switching events, we employ a counter (Tektronix FCA 3100) for faster measurements. A symmetric triangular wave current (larger than $I_\text{s}$) of frequency 10~Hz from a Keysight 33600A function generator is sent through the device. A $dc$ offset current is added to the triangular current. The offset makes the triangular current go from $0$ to $+I$ or $0$ to $-I$. The minimum and maximum value of the triangular current ensures the device switches from superconducting to the normal state and again returns to the superconducting state in each cycle of the triangular current. The counter measures the time interval between two events -- when the voltage drop across the device is zero at $I=0$ and when it reaches a non-zero pre-defined value because of switching from superconducting to normal state. From the measured time interval, we convert it to switching currents ($I_\text{s}^{+}$ and $I_\text{s}^{-}$). More details are provided in Supplementary Information, Section S2. 

\subsubsection*{Measurement of switching currents with $B$}

Instead of taking single $I-V$ curves and finding out $I_\text{s}^{+}$ and $I_\text{s}^{-}$, we take 100 switching events with a counter and average them to get $I_\text{s}^{+}$ and $I_\text{s}^{-}$ at each $B$. Here, $B$ is the slow axis for the measurement. A homemade electromagnet was used to get a very small $B$. The device, along with the electromagnet, was enclosed inside a cryoperm box to shield the device from outside stray fields. Extreme care has been taken to calibrate the actual fields inside the shielded cryoperm box. Details are provided in Supplementary Information, section S12.      

\section*{Data availability}

The data supporting the findings of this study are available from the corresponding authors upon reasonable request.

\bibliography{Josephson_diode}

\section*{Acknowledgments}

We thank Eli Zeldov, Abhay Pasupathy, Vibhor Singh, Pratap Raychaudhuri, Satyajit Banerjee, Sophie Gueron, H\'el\`ene Bouchiat, R. Vijay, Vladimir Krasnov, Subhajit Sinha, and Joydip Sarkar for helpful discussions and comments. We thank Amit P. Shah for his help in making SiN membranes for shadow masks. We acknowledge Department of Science and Technology (DST), Nanomission grant SR/NM/NS-45/2016, CORE grant CRG/2020/003836, and Department of Atomic Energy (DAE) of Government of India 12-R\&D-TFR-5.10-0100 for support. Work in Berlin was supported by Deutsche Forschungsgemeinschaft through CRC 183 (project C02) as well as TWISTGRAPH.

\section*{Author contributions}
S.G. and V.P. fabricated the devices. A.B., K. and A.D. helped in device fabrication. S.G. and A.B. did the measurements. D.A.J., R.K., and A.T. grew the BSCCO crystals. J.F.S. and F.v.O. developed the theoretical interpretation.
S.G., F.v.O.  and M.M.D. wrote the manuscript. All authors provided inputs to the manuscript. M.M.D. supervised the project.

\section*{Competing interests}
The authors declare no competing financial interests.

\clearpage



\setcounter{figure}{0}

\renewcommand{\figurename}{\textbf{Extended Data Fig.}}

 \makeatletter
\def\@fnsymbol#1{\ensuremath{\ifcase#1\or \dagger\or *\or \ddagger\or
   \mathsection\or \mathparagraph\or \|\or **\or \dagger\dagger
   \or \ddagger\ddagger \else\@ctrerr\fi}}
    \makeatother

\begin{figure*}[h]
\centering

\includegraphics[width=15.24cm]{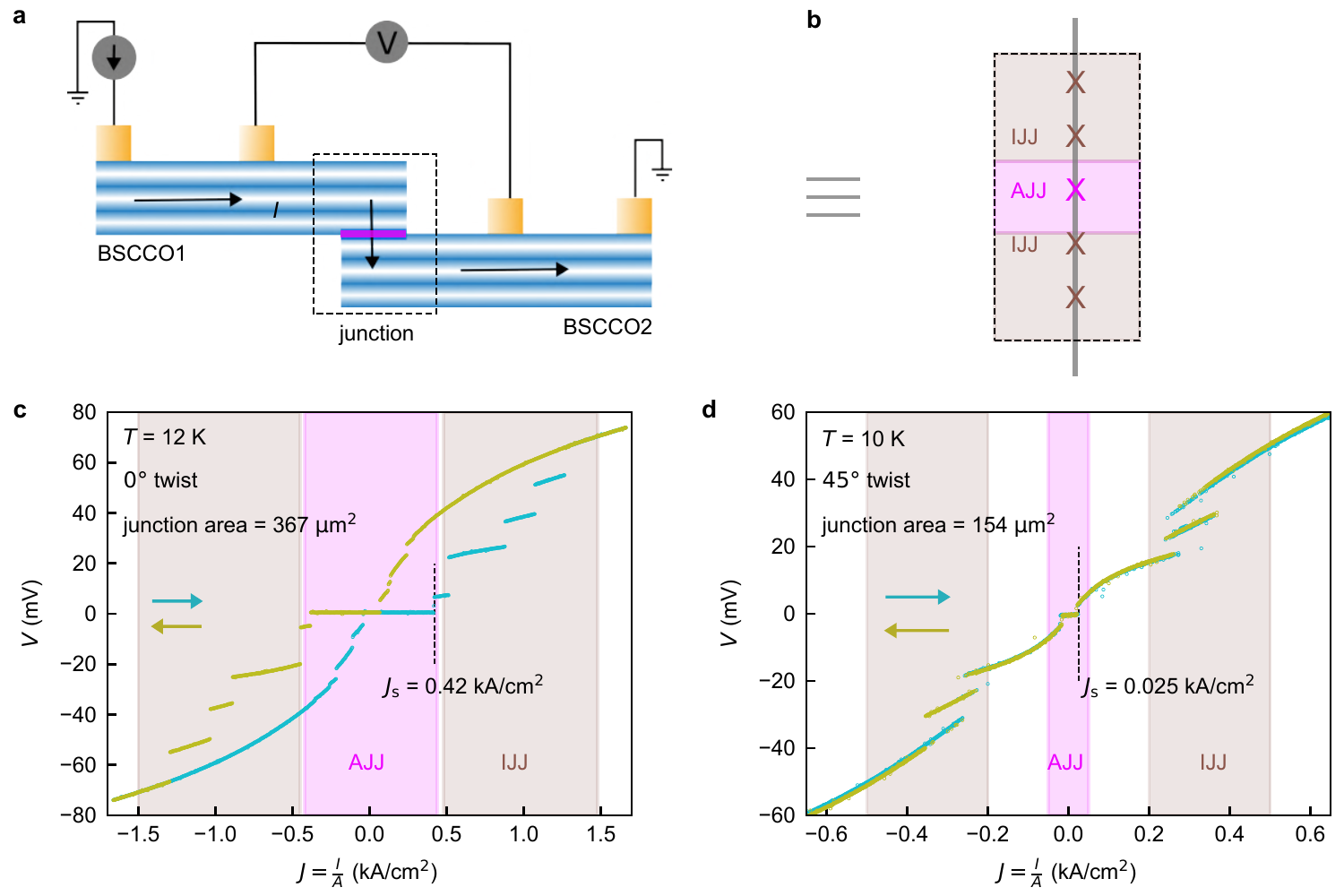}
\caption{\label{fig:fig5} \textbf{Modulation of switching current with twist angle.} (a) Cross-sectional schematic of twisted BSCCO junction. (b) Equivalent picture of the junction, marked by the dashed rectangle in (a). In the bulk of each BSCCO flake, there are series of intrinsic Josephson junctions (IJJs). At the twisted interface of two BSCCO flakes, an artificial Josephson junction (AJJ) is formed. (c), (d) \textit{dc} $I-V$ characteristics of 0\degree~and 45\degree~twisted junction at 12~K, and 10~K, respectively. Current in the $x$-axis is normalized with the area (A) of the junctions to get current densities ($J$). The magenta-shaded region originates at the interface due to AJJ, while the brown-shaded area is the contribution of IJJs coming from bulk. The switching current density of 45\degree~device is suppressed by a factor of 20 compared to 0\degree~twisted device. }

\end{figure*}

\clearpage


\begin{figure*}[h]
\centering

\includegraphics[width=17cm]{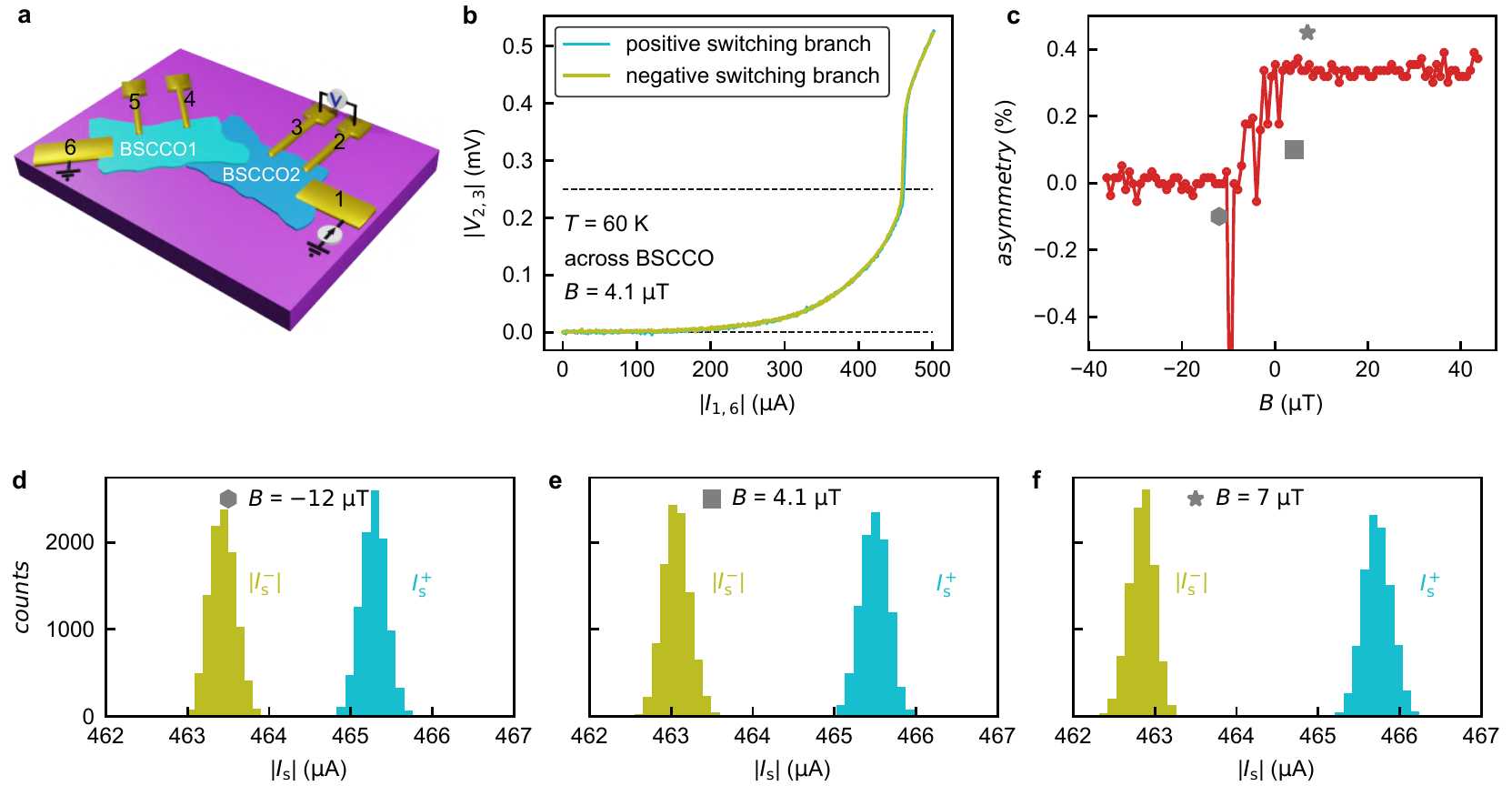}
\caption{\label{fig:fig6} \textbf{Negligible current asymmetry in  pristine BSCCO.} (a) Schematic of the measurement scheme. Current is sourced and collected with electrodes 1 and 6. Voltage drop across BSCCO is measured with electrodes 2 and 3. (b) $I-V$ characteristic across pristine BSCCO at 60~K. Negative bias switching branch is flipped to compare it with the positive bias switching branch. The dashed horizontal line ($\pm 0.25$ mV) indicates the triggering voltage used for obtaining switching currents with the help of a counter. (c) Switching current asymmetry for positive and negative bias with $B$, applied perpendicular to the device plane. Asymmetry across BSCCO is very small. (d), (e), (f) Distributions of switching currents ($I_\text{s}^{+}$ and $I_\text{s}^{-}$) at three different magnetic fields (marked by hexagon, square and star in (c)), $-12$, $4.1$, and $7$ \textmu T. Additionally we performed a second control experiment in pristine BSCCO where we restrict the bias current to flow only in the pristine BSCCO and not through the junction. The result is consistent (see Supplementary Information Fig. S14) and the asymmetry we observe is still very small ($< 1$ \%). }

\end{figure*}

\clearpage


\begin{figure*}[h]
\centering

\includegraphics[width=17cm]{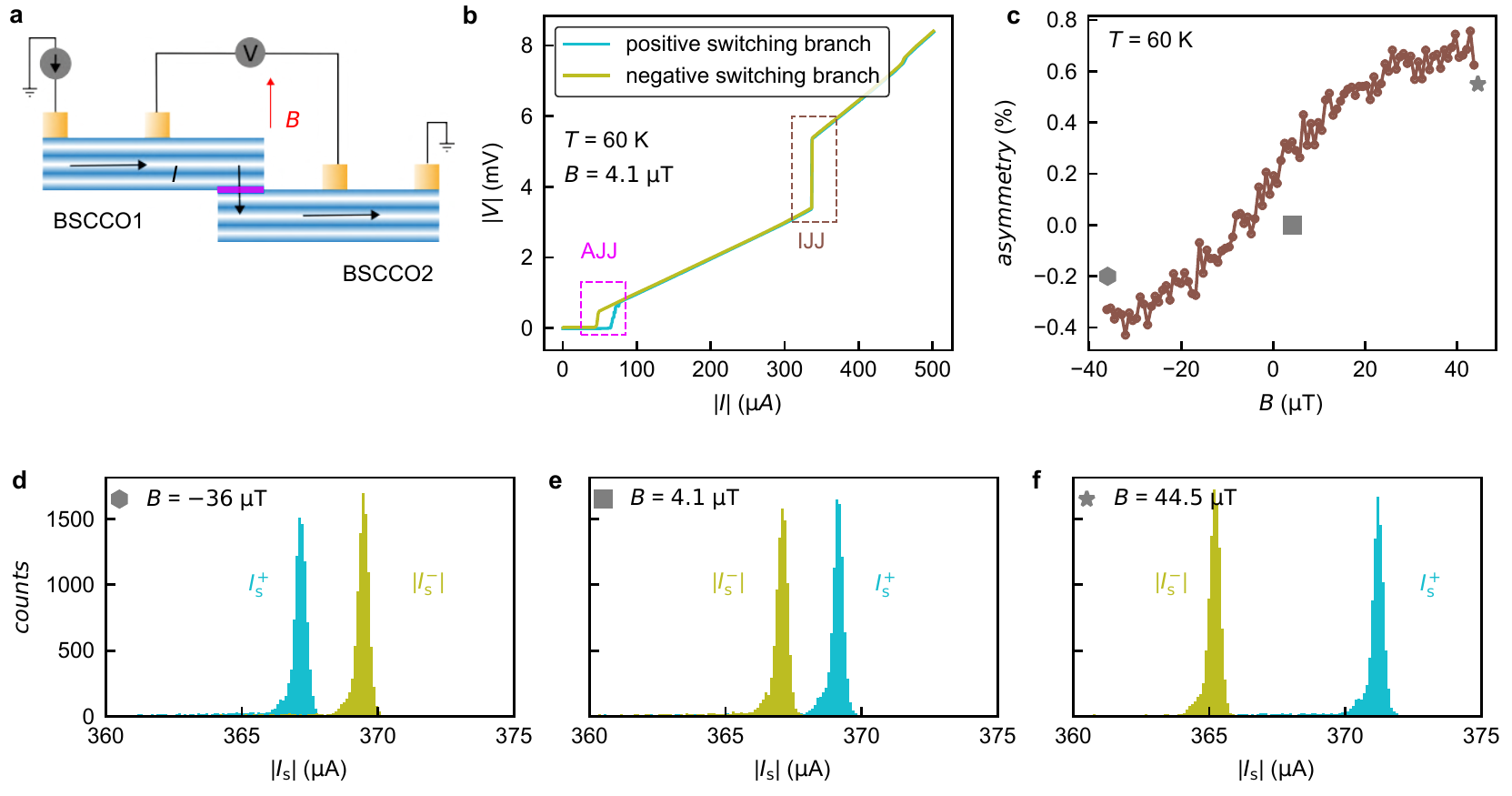}
\caption{\label{fig:fig7} \textbf{Negligible current asymmetry in  intrinsic Josephson junction (IJJ).} (a) Schematic of the measurement scheme. (b) $I-V$ characteristic across the junction at 60~K. The different colored curves are for positive and negative switching branches. AJJ switches at a smaller bias current, marked by a magenta dashed rectangle. IIJ switches at a higher bias current, marked by the brown dashed rectangle. (c) Asymmetry between positive ($I_\text{s}^{+}$) and negative ($I_\text{s}^{-}$) switching currents with $B$. At each $B$, 100 switching events were recorded and averaged to get $I_\text{s}^{+}$ and $I_\text{s}^{-}$. A trigger voltage of $\pm 3.6$ mV was used to get switching currents for IJJ. (d), (e), (f) Distributions of switching currents ($I_\text{s}^{+}$ and $I_\text{s}^{-}$) at three different magnetic fields (marked by hexagon, square and star in (c)), $-36$, $4.1$, and  $44.5$ \textmu T }

\end{figure*}

\clearpage


\begin{figure*}[h]
\centering

\includegraphics[width=17cm]{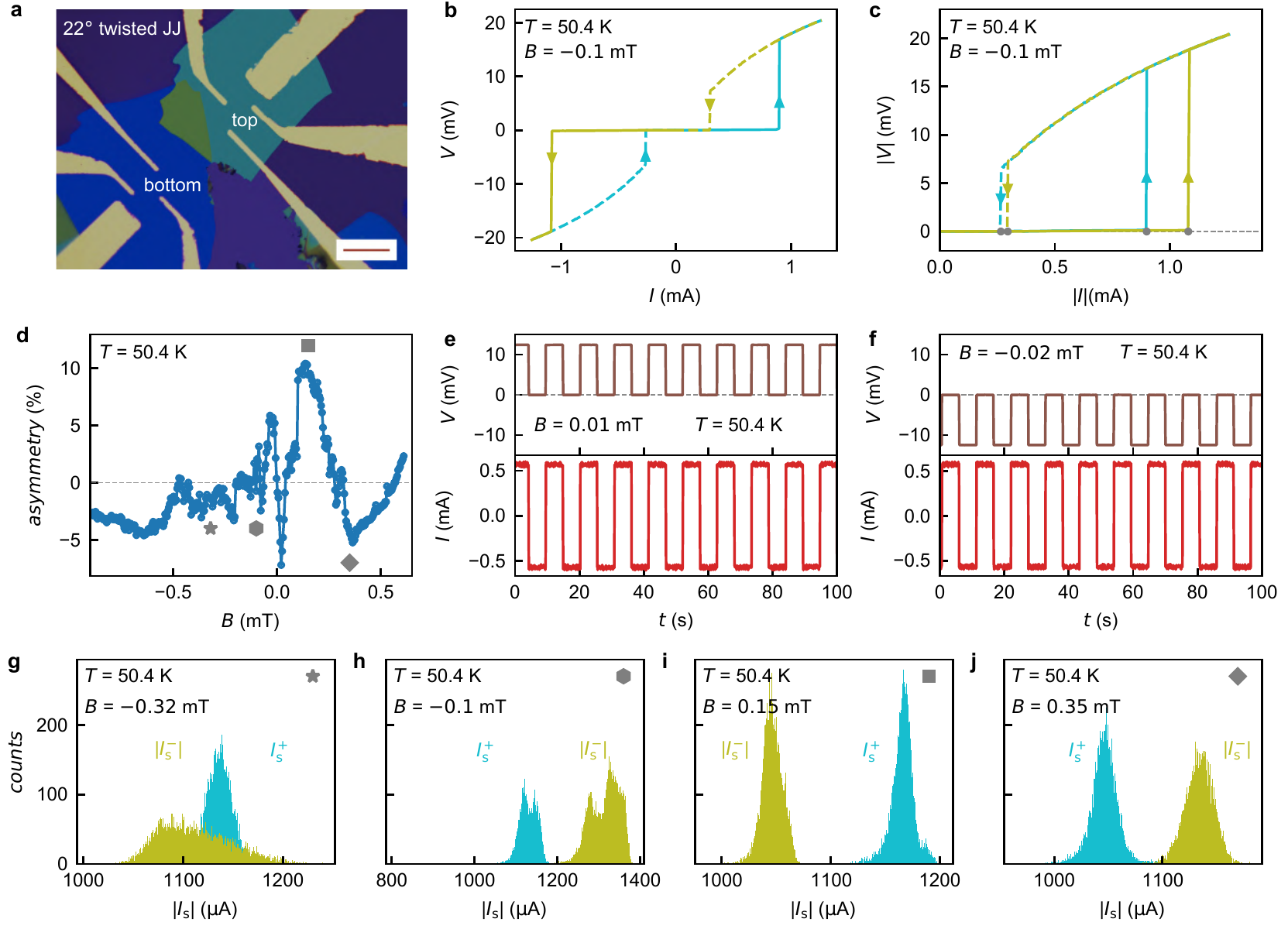}
\caption{ \label{fig:fig8}  \textbf{Superconducting diode behavior of 22\degree~twisted (D5) BSCCO Josephson junction.} (a) Optical micrograph image of 22\degree~twisted BSCCO JJ. The scale bar is 20 \textmu m. (b) $dc$ $I-V$ characteristic of the junction at 50.4~K and at $B = - 0.78$ mT. The two curves are for the up and down sweep of bias current. (c) Absolute values of $I$ and $V$ are plotted to clearly see the asymmetry. Here |$I_\text{s}^{-}$| is larger than $I_\text{s}^{+}$. (d) Variation of asymmetry factor $\alpha$ (defined in the text) with $B$ at 50.4~K. $B$ is applied perpendicular to the junction plane. For each $B$ value, 100 switching events were recorded and averaged to get $I_\text{s}^{+}$ and $I_\text{s}^{-}$. A trigger voltage of $\pm 0.2$ mV was used to get switching currents. The $\alpha$ is calculated from the average values of $I_\text{s}^{+}$ and $I_\text{s}^{-}$. Although $\alpha$ of the 22\degree~device varies with $B$ qualitatively like the 45\degree~and 0\degree~devices, there are more complex details to it. The measurement has been done in cryogenic setup 2 with a superconducting magnet. An offset value of 1 mT has been subtracted due to the trapped flux in the superconducting magnet (see Supplementary Information, section S13 for details). (e),(f) Half-wave rectification response of the junction at 50.4~K for two different $B$ values. (g),(h), (i), and (j) Distributions of the switching currents ($I_\text{s}^{+}$ and $I_\text{s}^{-}$) at different $B$ values. $10^4$ switching events were recorded to get the histograms.} 
\end{figure*}

\begin{figure*}
\centering

\includegraphics[width=12.7cm]{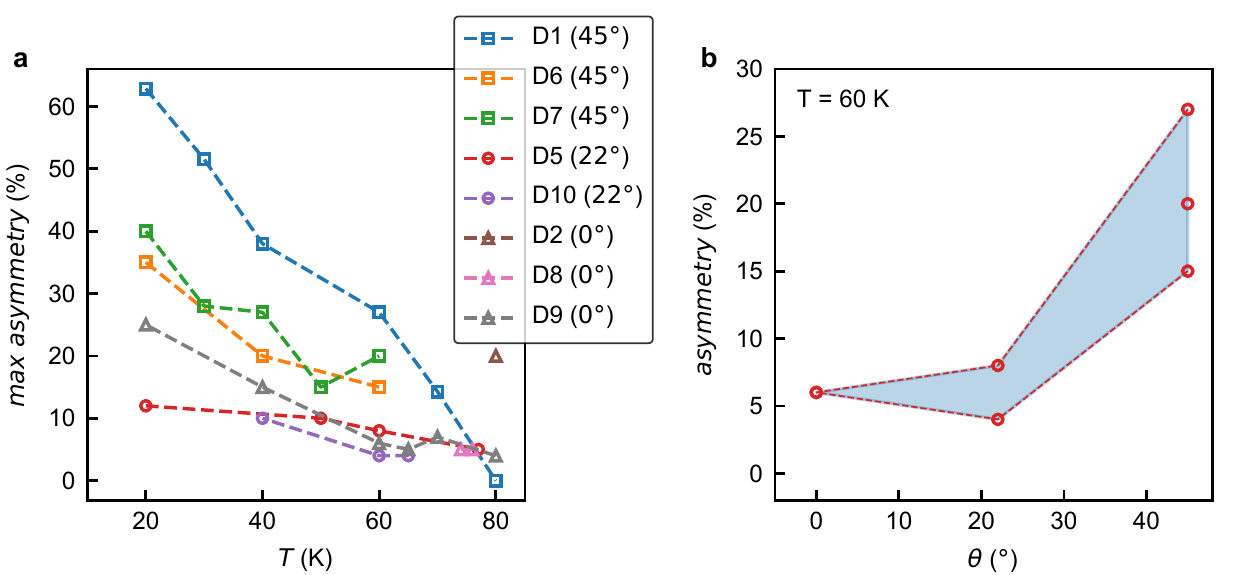}
\caption{ \label{fig:fig9}  \textbf{Temperature and twist angle dependent asymmetry.} We have measured multiple devices at three specific twist angles ($0\degree$, $22\degree$ and $45\degree$). (a) shows temperature dependence of maximum switching current asymmetry value for devices with different twist angles. For all the devices, the asymmetry increases at lower temperatures. We do not have data points at lower temperatures for all the devices with $0\degree$ twist angle; this is mainly to avoid damage to the devices because of large switching currents at lower temperatures (switching current is strongly dependent on twist angle). (b) shows twist angle dependence of asymmetry value at 60~K. The asymmetry value is maximum for $45\degree$ twisted devices and decreases with a decrease in twist angle.} 
\end{figure*}

\begin{figure*}
\centering

\includegraphics[width=8.9cm]{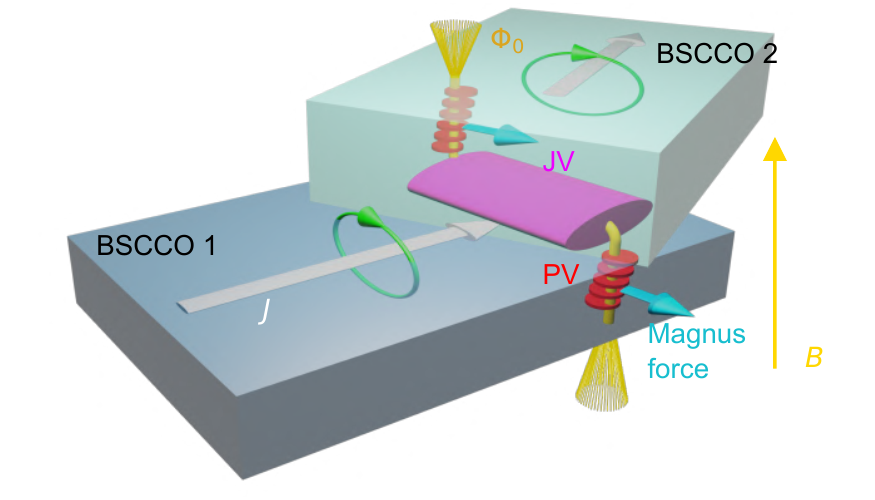}
\caption{ \label{fig:fig10}  \textbf{Mechanism of diode effect.} Schematic of a twisted BSCCO junction with possible vortex configuration. Out-of-plane $B$ induces stacks of pancake vortices (PV) which can be pinned at different locations in top and bottom flake and connect through a Josephson vortex (JV). The difference in vortex configurations along with the in-plane field of JV result in a spatially nonuniform gauge-invariant phase difference across the AJJ and thus Fraunhofer-like interference. Changes in the vortex configurations due to the Magnus force exerted by the bias current ($J$) will differ depending on bias direction, resulting in diode behavior. We estimate that self-fields of the bias current (green loops) contribute only weakly to the asymmetry.}
\end{figure*}


\clearpage

\begin{center}
\textbf{\LARGE Supplementary Information: High-temperature Josephson diode}
\end{center}

\renewcommand{\figurename}{\textbf{Fig.}}

\setcounter{equation}{0}
\setcounter{figure}{0}
\setcounter{table}{0}
\setcounter{page}{1}

\renewcommand{\thesection}{S\arabic{section}}
\renewcommand{\theequation}{S\arabic{equation}}
\renewcommand{\thefigure}{S\arabic{figure}}
\renewcommand{\thepage}{S\arabic{page}}

 \makeatletter
\def\@fnsymbol#1{\ensuremath{\ifcase#1\or \dagger\or *\or \ddagger\or
   \mathsection\or \mathparagraph\or \|\or **\or \dagger\dagger
   \or \ddagger\ddagger \else\@ctrerr\fi}}
    \makeatother


\section{Cryogenic exfoliation to make twisted BSCCO JJ}

\begin{figure*}[h]
\centering

\includegraphics[width=16cm]{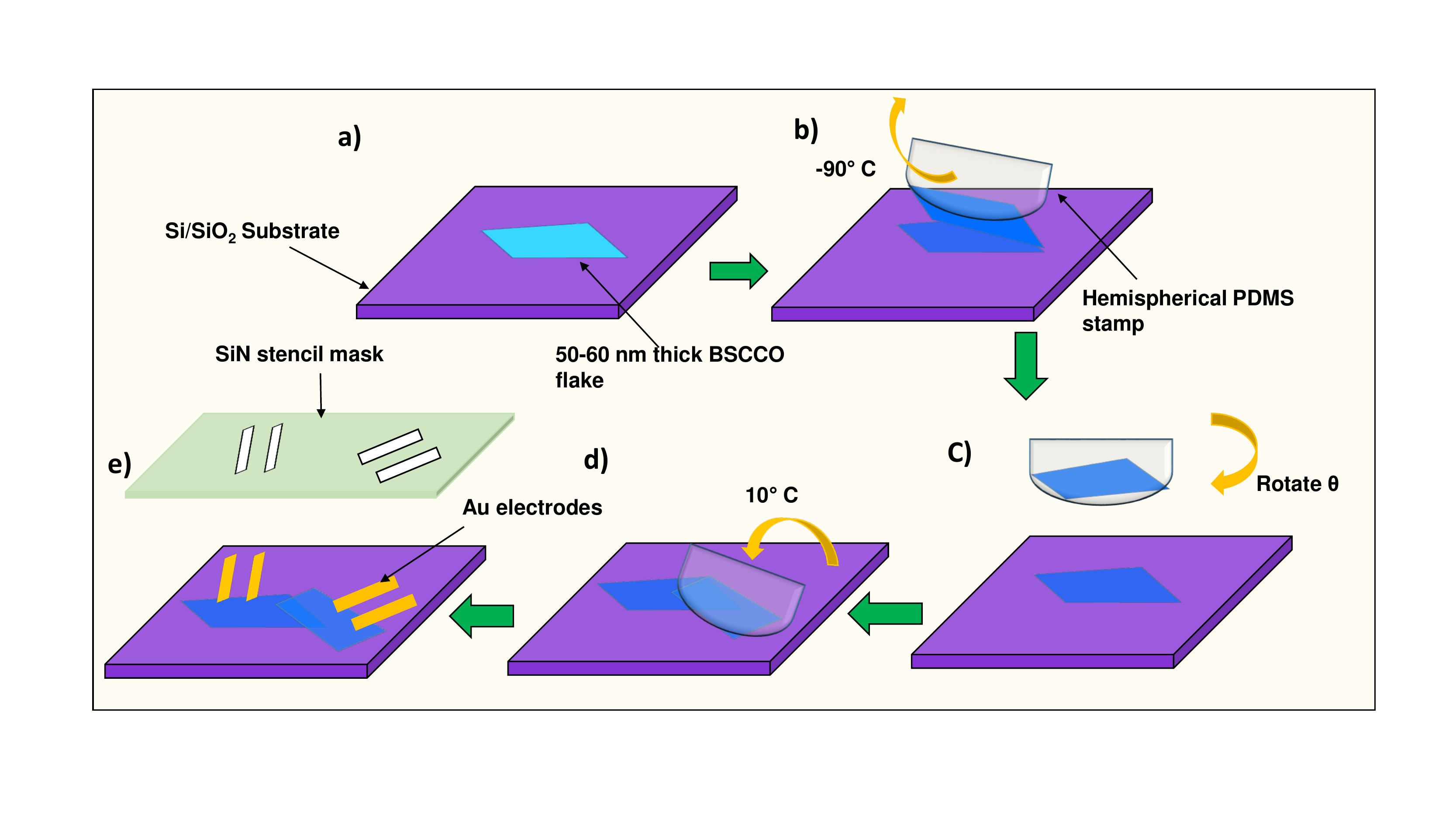}
\captionsetup{font=footnotesize}
\caption{ \label{fig:figS1}  \textbf{Twisted BSCCO device fabrication by cryogenic exfoliation} (a) 50-60 nm thick BSCCO flake exfoliated on Si/SiO$_2$. (b) Re-exfoliation of the thick BSCCO flake with a hemispherical PDMS stamp at low temperature ($-90$\degree~C) on a cold stage. (c) The hemispherical PDMS stamp is quickly rotated by an angle and stacked onto the bottom flake. (d) The cold stage is warmed to 10\degree~C, and PDMS is slowly removed (e) The Gold (Au) electrodes are deposited through a SiN stencil mask on top of both flakes.}

\end{figure*}

Some 2D materials are chemically unstable in the atmosphere. Therefore, the device fabrication of those 2D materials is challenging. In this work, we developed cryogenic exfoliation, pick-up, and transfer methods for 2D materials following Zhao \textit{et al.} \cite{zhao_emergent_2021}. BSCCO degrades with time; hence we used hemispherical PDMS dry transfer techniques for the twisted device fabrication in the glove box environment. The hemispherical PDMS stamp is made using Sigma Aldrich Sylgard 184 silicon elastomer base and curing agent mixture with a ratio of 10:1. After that, we cut Gel-Pak (part no- 40x40-0170-X4) PDMS  into cylindrical pieces with a diameter of $\sim~2$ mm and put the viscous mixture of Sylgard 184 on those pieces, then baked at 100\degree~C for 15 min. For the exfoliation of BSCCO crystal, we used a Si/SiO$_2$ substrate, which was cleaned by O$_2$ plasma for 90 s and heated at 150\degree~C overnight. Then we exfoliate BSSCO (Bi-2212) crystal by the mechanical scotch tape method and transfer the exfoliated BSCCO flake onto the pre-cleaned Si/SiO$_2$ substrate. Then we put the substrate on the cryogenic stage and search for thick ($\sim$ 40-60 nm), flat, and uniform BSCCO flake through the optical microscope. Once we identify the desired flake, we cool down the cryogenic setup using liquid nitrogen. When the temperature reaches $-25\degree$~C, we attach a hemispherical PDMS stamp to the BSCCO flake and wait for the temperature to reach $\sim-90\degree$~C. At this temperature, the PDMS goes through the glass transition state. Because of that, the PDMS meniscus automatically detaches from the BSCCO flake. During this detachment, some part of the flake cleaves and attaches to the PDMS stamp. Then we quickly rotate the PDMS stamp at an angle (with an accuracy of $\pm 0.5\degree$) and drop the PDMS-attached BSCCO flake on top of the bottom flake, as shown in Fig.~\ref{fig:figS1}. Then, we warm the cold stage to $10\degree$~C and slowly remove the hemispherical PDMS stamp. Finally, we align a SiN mask with the flakes and deposit 70 nm thick gold (Au) metal electrodes by e-beam evaporation. After completing device fabrication, we immediately load the device in a high vacuum cryostat for measurement.

\section{Protocol for measuring switching currents with counter}

\begin{figure*}[h]
\centering

\includegraphics[width=17cm]{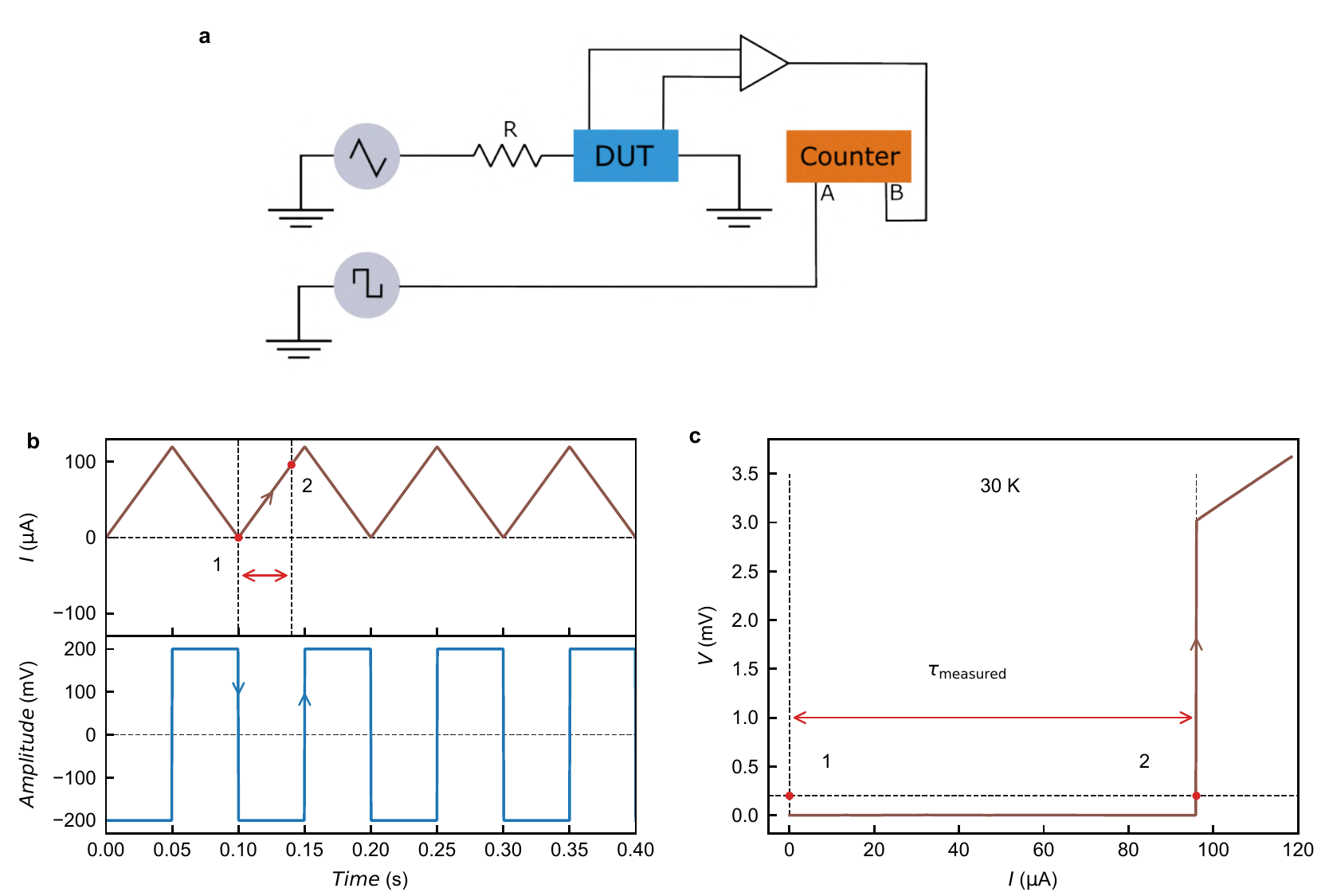}
\captionsetup{font=footnotesize}
\caption{ \label{fig:figS2}  \textbf{Switching statistics measurement using a frequency counter} (a) Circuit diagram for the measurement of the switching statistics. The square wave signal is directly connected to one of the ports of the counter. A triangular voltage source is connected to the device under test (DUT) through a series resistor, and the voltage drop across the junction is connected to the other port of the counter after amplification. (b) Triggering action of the counter for the two events. The square signal is a reference for triggering the first event when the system is in the superconducting state. (c) $dc$ $I-V$ characteristic of the 45\degree~twisted device (D1) at 30~K. The switching of the device from superconducting to normal state is identified when the voltage drop reaches 0.2 mV. This is the second trigger point for the counter. From the measured time interval between events 1 and 2, we calculate the switching currents. The trigger voltage for a particular device depends on its normal state resistance and varies from device to device.} 
\end{figure*}

Tektronix FCA 3100 frequency counter and Keysight 33600A waveform generator are used to measure the statistics of switching currents. This measurement protocol is much faster than taking several $dc$ $I-V$ characteristics and then finding out the distributions. The counter measures the time interval between two events. From the measured time interval, we then convert it to the switching current. We send a symmetric triangular current wave of frequency 10~Hz and a magnitude larger than the switching currents, through the device. A $dc$ offset current with appropriate magnitude is added to the triangular current. For example, in Fig.~\ref{fig:figS2}b the triangular current goes from 0 to the positive maximum value after adding an offset. The role of the $dc$ offset current is to ensure that at each cycle of the triangular current, the system goes from superconducting to the normal state. We feed a square wave reference signal with arbitrary amplitude but with the same frequency of 10~Hz to one of the ports (port A in Fig.~\ref{fig:figS2}a) of the counter. A 90\degree~phase difference is added to the triangular current with respect to the square reference signal. As a result, the triangular current has a rising edge when the reference square wave has a falling edge, as shown in Fig.~\ref{fig:figS2}a. We configure the counter such that it triggers at this point and starts measuring the time. At this point, the system is in the superconducting state. This is the first event. The voltage drop (due to the triangular current wave) across the device is fed to the other port of the counter (port B in Fig.~\ref{fig:figS2}a). The second event triggers when the voltage drop across the device (fed to port B of the counter) reaches a predefined value, as indicated by the dashed horizontal line in Fig.~\ref{fig:figS2}c. This predefined trigger value is not fixed and varies from device to device depending on the normal state resistance of the devices. At this point, the system has switched to the normal state. We calculate the switching currents from this measured time interval between the two events. At each cycle, the counter measures the time interval. We take $10^4$ such counts for the distributions.  

\section{Superconducting transition temperature ($T_\text{c}$) of artificially made JJ and pristine BSCCO}

\begin{figure*}[h]
\centering

\includegraphics[width=17 cm]{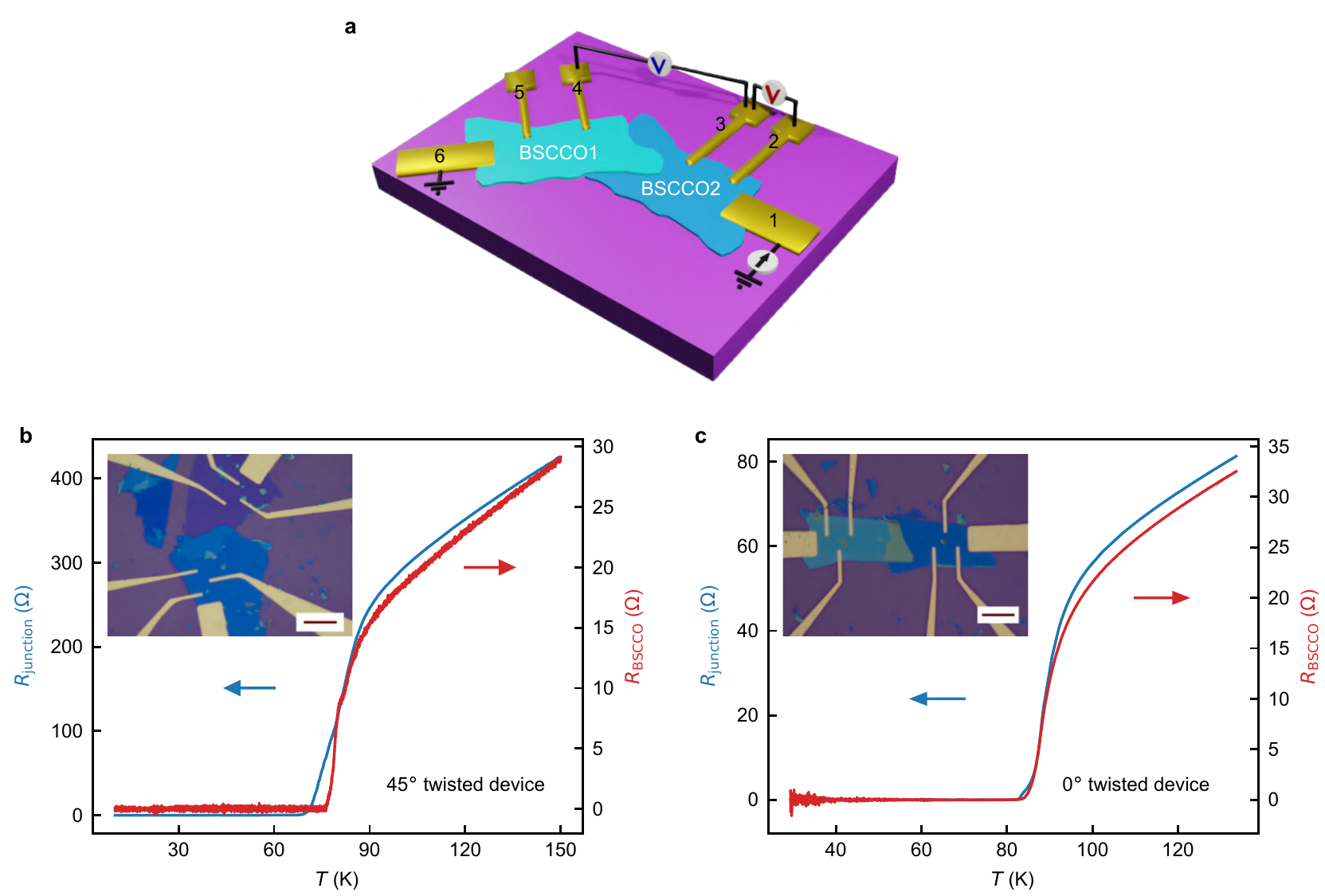}
\captionsetup{font=footnotesize}
\caption{ \label{fig:figS3}  \textbf{$R$ vs $T$ response of 45\degree~(D1) and 0\degree~(D2) twisted device.} (a) Schematic of four probe $R$ vs $T$ measurements. We simultaneously measure the junction (by blue voltmeter) and pristine BSCCO (by red voltmeter) to ensure no degradation during fabrication steps. (b) $R$ vs $T$ response of the 45\degree~twisted device (D1). The blue and red curves are the response from the twisted junction and pristine BSCCO, respectively. The inset shows an optical micrograph of the device. The scale bar is 20 \textmu m. (c) $R$ vs $T$ response of the 0\degree~twisted device (D2). The blue and red curves are the response from the twisted junction and pristine BSCCO, respectively. The inset shows an optical micrograph of the device. The scale bar is 20 \textmu m.}

\end{figure*}

We measure four probe $R$ vs $T$ across the artificial junction and pristine BSCCO for all the devices. Fig.\ref{fig:figS3}a and Fig.\ref{fig:figS3}b show the $R$ vs $T$ of the 45\degree~(D1) and 0\degree~(D2) devices of the main manuscript. The superconducting transition temperatures ($T_\text{c}$) of the junction and that of the pristine BSCCO flake are close for both 45\degree~and 0\degree~devices.

\section{Comparison of maximum current asymmetry and operating temperature with earlier reports}

A variety of systems have been used to probe the superconducting diode effect. Here we tabulate some of the key results showing the reported values of maximum asymmetry and the operating temperatures.

\captionof{table}{Literature review of SDE/JDE observed in a large variety of systems and comparison of maximum asymmetry and operating temperature with our device.}

\begin{tabularx}{1\textwidth} { 
    | >{\raggedright\arraybackslash}X 
    | >{\raggedright\arraybackslash}X
    | >{\raggedright\arraybackslash}X
    | >{\raggedright\arraybackslash}X | }

    \hline
    
    Author & System & Maximum asymmetry ($\frac{\Delta I_c}{I_{c_+}+|I_{c_-}|}\times 100 \% $) & Maximum operational temperature \\
    
    \hline
    
    Ando et al. \cite{ando_observation_2020}
	& Nb/V/Ta & 1.8 \% at 4.2 K & 4.35 K\\ 
	
	\hline
	
	Narita et al. \cite{narita_field-free_2022}
	&  Nb/V/Co/V/Ta & 3.9\% at 1.9 K at -0.04 T & 2.4 K \\ 
	
	\hline
	
	Wu et al. \cite{wu_field-free_2022} & NbSe$_2$/Nb$_3$Br$_8$/ NbSe$_2$ & 14.2\% at 35 mT, 20 mK  & 3.86 K\\
	
	\hline
	
	Bauriedl et al. \cite{bauriedl_supercurrent_2022} & hBN/NbSe$_2$/hBN & 35 \% at 35 mT at 1K & 4.3 K\\
	
	\hline
	
	Lin et al. \cite{lin_zero-field_2022} & MATTLG  & 55\%   at T= 20 mK & 1.6 K\\
	
	\hline
	
	Pal et al. \cite{pal_josephson_2022} & Nb/NiTe$_2$/Nb & 30\% at B$_\mathrm{y}$ = 12 mT at 60 mK & 4 K \\ 
	
	\hline
	
	Shin et al. \cite{shin_magnetic_2021} & CrPS$_4$/NbSe$_2$/CrPS$_4$ & 7-9 \% at B=10 mT at 1.55 K & 4.5 K\\
	
	\hline
	
	\textcolor{red}{Our device} & \textcolor{red}{twisted BSCCO JJ} & 		\textcolor{red}{60 \% at 20 K} & \textcolor{red}{80 K}\\
	
	\hline

\end{tabularx}


\section{Tunability of switching distributions with magnetic field}

\begin{figure*}[h]
\centering

\includegraphics[width=12.7cm]{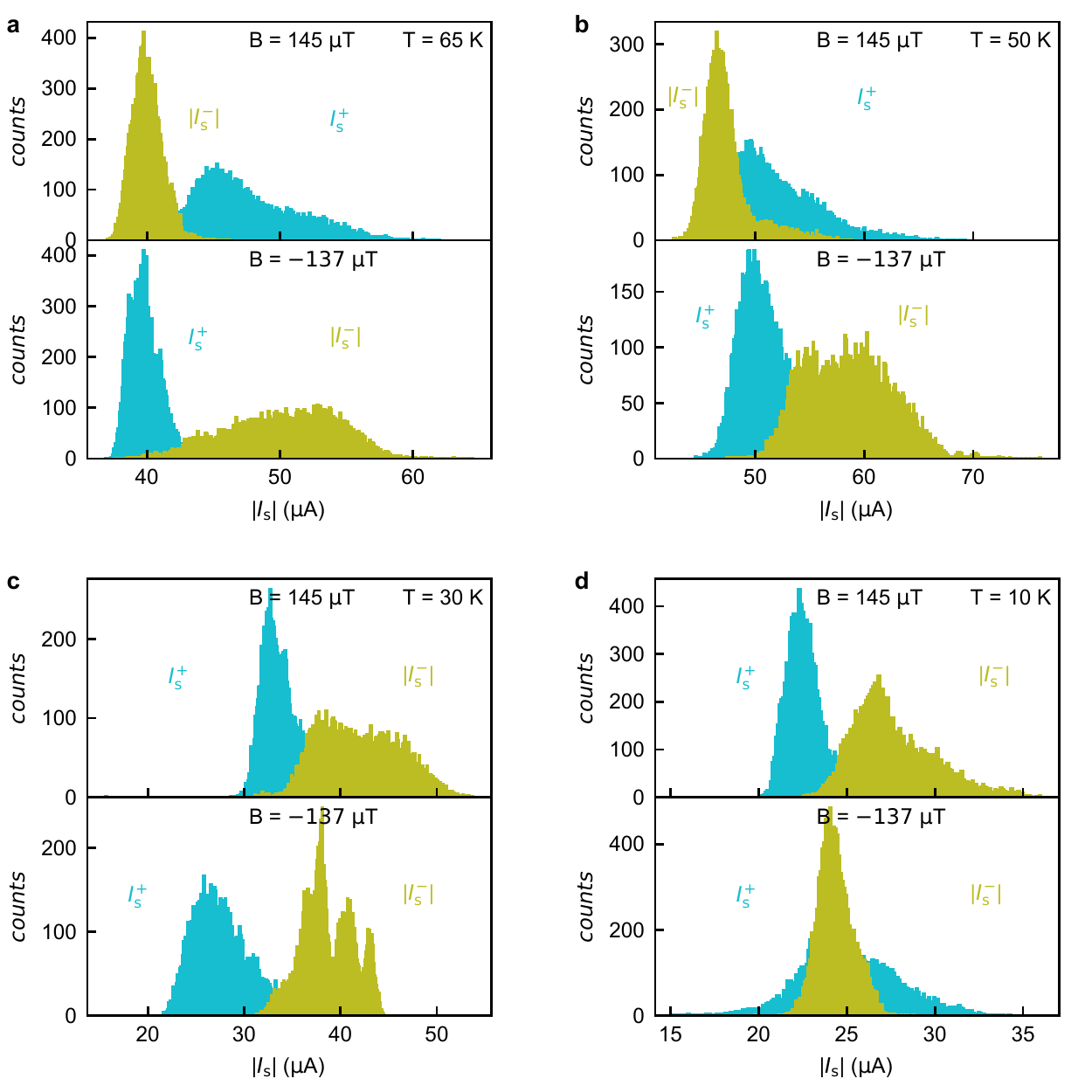}
\captionsetup{font=footnotesize}
\caption{ \label{fig:figS4} \textbf{Switching distributions with $B$ at different temperatures for 45\degree~twisted (D1) BSCCO JJ.} The histograms are the result of $10^4$ switching events for both positive and negative switching currents. The spread of the distributions for $I_\text{s}^{+}$ and $I_\text{s}^{-}$ are different, and flips sign with the flipping sign of $B$.}

\end{figure*}

As shown in the main manuscript, the asymmetry $\alpha$ is antisymmetric in $B$. The antisymmetric nature is also reflected in the distribution spread of switching currents $I_\text{s}^{+}$ and $I_\text{s}^{-}$. In the main manuscript, we have shown the spread of the distribution changes with a change in sign of $B$ near maximum asymmetry $\alpha$. We observe similar behavior away from maximum $\alpha$ as well. Fig.~\ref{fig:figS4}a-d show the switching current distributions at higher $B$ values at different temperatures; there is a very small asymmetry between the magnitude of $I_\text{s}^{+}$ and $I_\text{s}^{-}$ for the 45\degree~twisted device. At all temperatures, the spread of the distributions of $I_\text{s}^{+}$ and $I_\text{s}^{-}$ flips with the change in sign of $B$. The substructures in the switching distributions are a consequence of the multistability of vortex configurations in these twisted devices which is discussed in \ref{Sec:distribution}.

\section{ Half-wave rectification at various temperatures by 45\degree~twisted device (D1)}

\begin{figure*}[h]
\centering

\includegraphics[width=17cm]{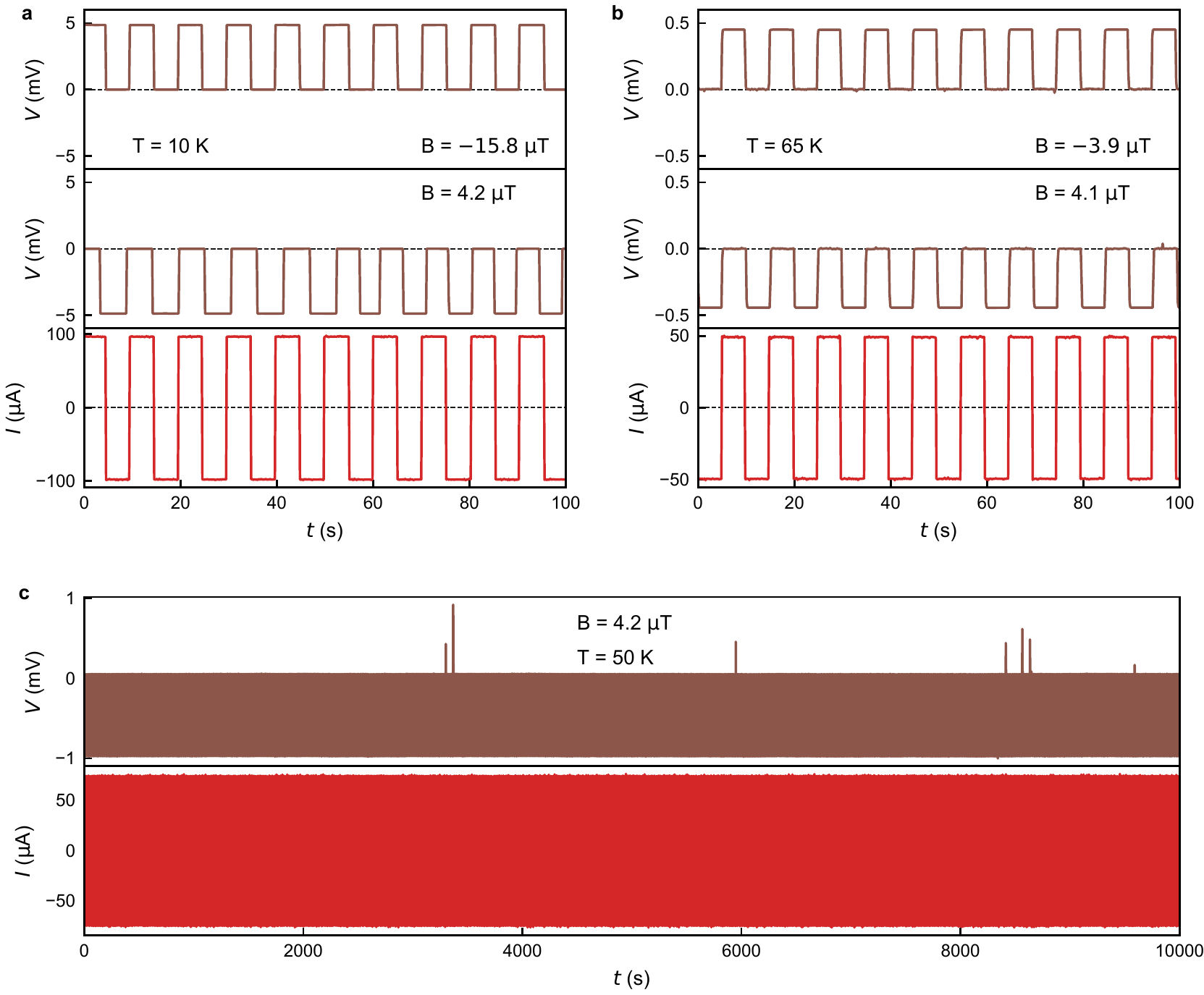}
\captionsetup{font=footnotesize}
\caption{ \label{fig:figS5} \textbf{Half-wave rectification at different temperatures.} (a), (b) Rectification action by the 45\degree~twisted junction (D1) at 10 K and at 65 K. (c) Rectification of 1000 cycles at 50 K. }

\end{figure*}

Fig.~\ref{fig:figS5} shows the half-wave rectification of a square wave current by the 45\degree~twisted junction (D1) at different temperatures. The rectification nature is tunable with applied magnetic field $B$, as shown in Fig.~\ref{fig:figS5}a, and b. In Fig.~\ref{fig:figS5}c, we show half-wave rectification of 1000 cycles at 50~K. Out of 1000 cycles, we see 7 events where the switching happened in the positive cycle.  

\section{Asymmetric $I-V$ for other of $0\degree$, $22\degree$ and $45\degree$ twisted JJs}

\begin{figure*}[h]
\centering

\includegraphics[width=17cm]{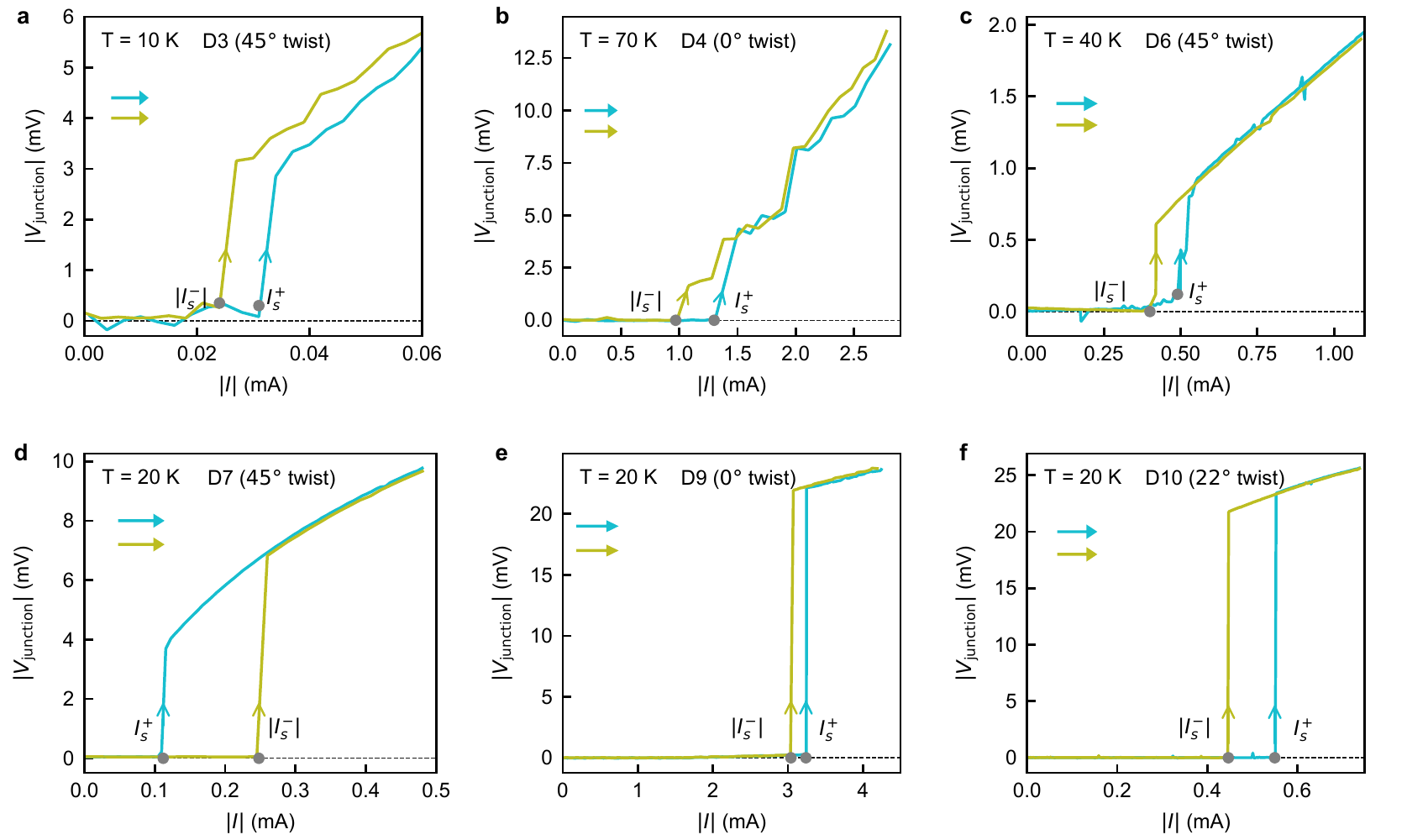}
\captionsetup{font=footnotesize}
\caption{ \label{fig:figS6}  \textbf{Switching current asymmetry for other $45\degree$, $22\degree$ and $0\degree$ twisted devices.} Asymmetry of switching currents for the different twisted devices. Negative branches of $I-V$ characteristics are flipped and plotted on an absolute scale to clearly see the difference between $I_\text{s}^{+}$ and $I_\text{s}^{-}$. Detailed field-dependent asymmetry and switching distributions were not done for D3 and D4. For detailed field-dependent asymmetry and distributions for switching currents, we used appropriate trigger voltages (0.2 mV in D6, 0.4 mV in D7, 2.5 mV in D9, and 2 mV in D10) in counter measurement.} 
\end{figure*}

We have fabricated multiple devices with twist angles $0\degree$, $22\degree$, and $45\degree$. $dc$ $I-V$ characteristics and the asymmetry in the switching current of these devices are shown in Fig.~\ref{fig:figS6}. We have done a detailed field dependence study on these devices except D3 and D4. The temperature and angle dependence of the maximum asymmetry of all these devices are shown in Extended Data Fig. 5 in the main manuscript.

\section{Control experiment in a BSCCO step edge device}

\begin{figure*}[h]
\centering

\includegraphics[width=17cm]{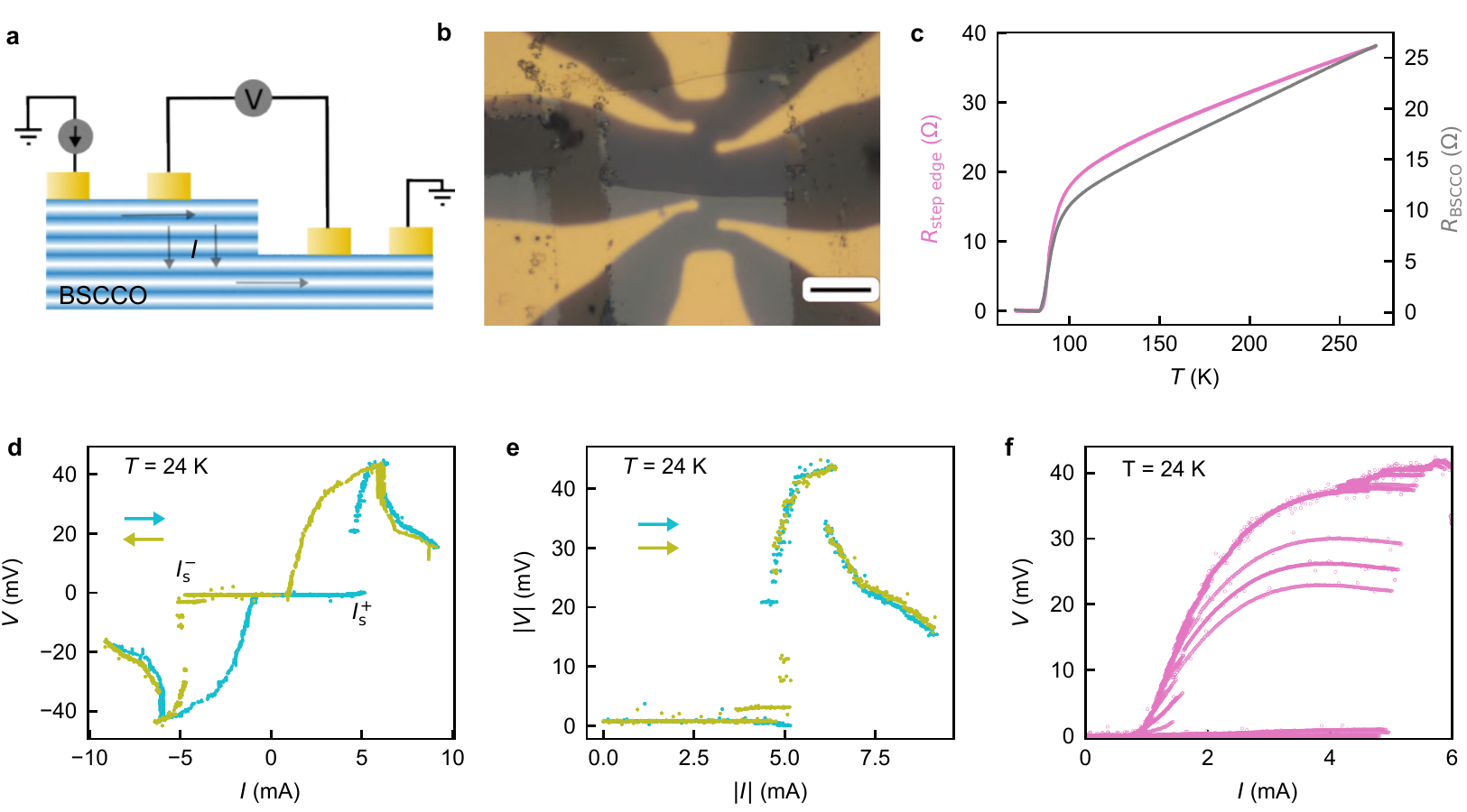}
\captionsetup{font=footnotesize}
\caption{ \label{fig:figS7}  \textbf{Control measurement in a BSCCO step edge device.} (a) Cross-sectional schematic of the step edge device. (b) Optical micrograph of the device. The scale bar shown in the figure is 20 \textmu m. Step edge appears spontaneously during the exfoliation process. (c) \textit{R} vs \textit{T} response across step edge and pristine BSCCO. (d) \textit{dc} $I-V$ characteristics across the step edge at 24 K. Different colors are used for two sweep directions of bias current, as indicated by the arrows. (e) The same \textit{dc} $I-V$ characteristics of (d), but the negative bias switching branch is flipped to compare with the positive bias switching branch. No significant asymmetry is there between the switching currents for positive and negative bias. (f) Branching in \textit{dc} $I-V$ response across the step edge. The branching appears due to the IJJs at the step edge.}

\end{figure*}

As an additional control test, we study the response of IJJs where there is no artificial junction. We use a natural step edge BSCCO flake to do this. The natural step edge appears stochastically during the exfoliation process. Gold electrodes are deposited to make electrical contacts to the flake using the shadow mask evaporation technique. Fig.~\ref{fig:figS7}a, b show cross-sectional schematic and optical micrograph of the step edge device. As shown in Fig.~\ref{fig:figS7}a, the current path will involve IJJs along the c-axis at the step edge. Fig.~\ref{fig:figS7}c shows $R$ vs $T$ across the natural step edge and pristine BSCCO. In Fig.~\ref{fig:figS7}d, \textit{dc} $I-V$ characteristics of the step edge device are shown for two opposite sweep directions of bias current. The negative bias switching branch is plotted by flipping its sign along with the positive bias switching branch in Fig.~\ref{fig:figS7}e. No significant difference is seen between $I_\text{s}^{+}$ and $I_\text{s}^{-}$, as was the case for IIJs of the twisted devices in the main manuscript. Fig.~\ref{fig:figS7}f shows the branching in the \textit{dc} $I-V$ characteristics at 24~K, a signature for c-axis transport involving IJJs \cite{kleiner_intrinsic_1992}.

\section{Control experiment in a BSCCO step edge device with B}

\begin{figure*}[h]
\centering

\includegraphics[width=12.7cm]{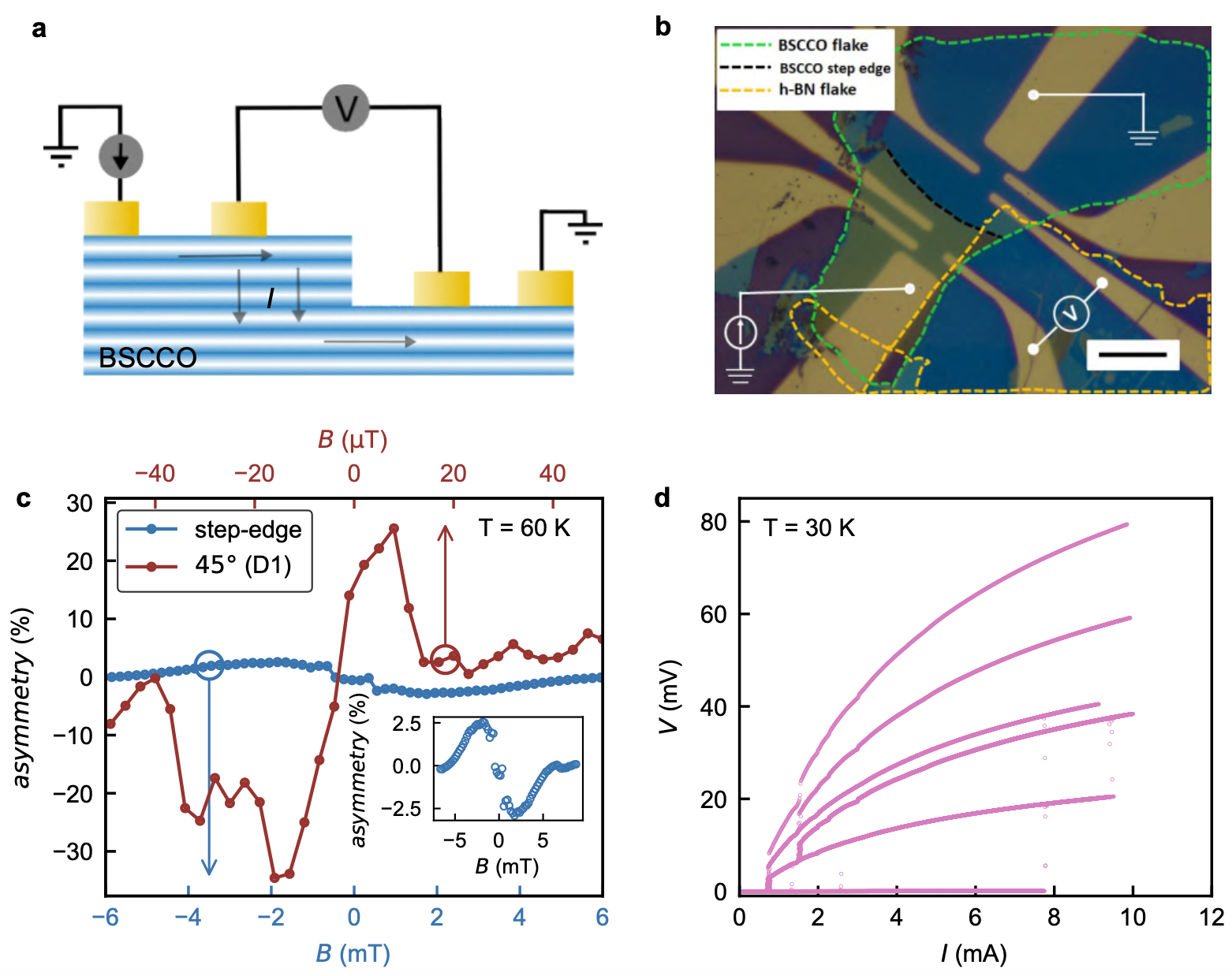}
\captionsetup{font=footnotesize}
\caption{ \label{fig:step_edge_2}  \textbf{Control step-edge BSCCO device with edges covered with hBN.} (a) Schematic of the step-edge device. (b) Optical micrograph of the step-edge device. Two edges of the thick part of the flake are covered with hBN, indicated by the dashed yellow line. Scale bar is 20 \textmu m. The electrodes for current biasing and voltage measurement across the step edge are indicated. The thickness of the thick part of the flake is 75 nm and the thin part is 30 nm. (c) Out-of-plane $B$-dependent switching current asymmetry for the step-edge device (blue data points) along with the asymmetry in a 45\degree (brown data points) twisted device (D1) at the same temperature of 60~K. At each $B$ 100 switching events we recorded with counter to get average switching currents. Note the different magnetic field scales for the step-edge (lower axis) device and twisted device (upper axis) as indicated by the arrows. Inset shows zoomed-in plot of asymmetry with $B$ for the step edge device showing small asymmetry. (d) Branching in the $dc$ $I-V$ characteristics across step-edge showing \textit{c}-axis transport that involves intrinsic JJs.}

\end{figure*}

We made another step-edge device to measure the $B$ dependence of the switching current asymmetry. We have covered the thick part of the step-edge with h-BN (Fig.~\ref{fig:step_edge_2}b) to make sure the current injecting electrode does not touch the side walls. This will ensure that the current is injected from the top surface of the thicker part of the flake to the thinner part. Fig.~\ref{fig:step_edge_2}d shows branching in the $dc$ $I-V$ characteristics across the step-edge; this is the signature of the \textit{c}-axis transport involving intrinsic Josephson junctions in the BSCCO. The out-of-plane $B$ dependence of the smallest switching current asymmetry of the step-edge device is plotted in Fig.~\ref{fig:step_edge_2}c along with the asymmetry of a $45\degree$ twisted device (D1) at the same temperature (60~K). Note that the $B$ range for both of the devices are very different and plotted on two scales (top for $45\degree$ twisted device and bottom for the step-edge device) as indicated by the arrows. We observe a small asymmetry for the step-edge device compared to the twisted device. Inset of Fig.~\ref{fig:step_edge_2}c shows a zoomed-in image of the same step-edge device. The small asymmetry in step-edge devices can be due to local doping variation and inhomogeneity in BSCCO that is well known. We want to highlight the point that twist angle-dependent asymmetry (Extended Data Fig. 5) is central to our observation of JDE in twisted BSCCO devices.

\section{Checking asymmetry in the normal state of the AJJ }

\begin{figure*}[h]
\centering

\includegraphics[width=9cm]{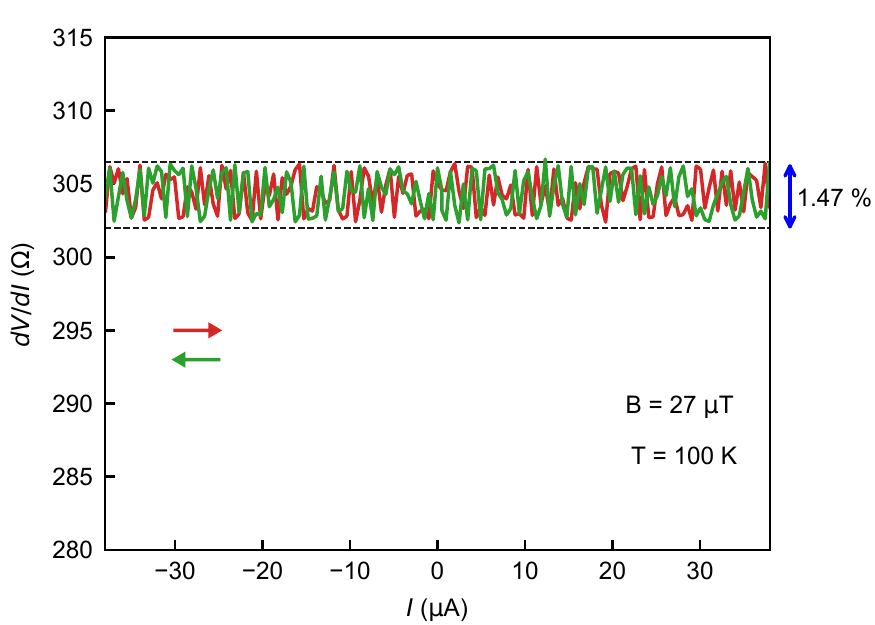}
\captionsetup{font=footnotesize}
\caption{ \label{fig:figS9}  \textbf{Normal state $\frac{dV}{dI}$ response of the 45\degree~twisted junction (D1) at 100 K for an applied magnetic field of 27 \textmu T.}}

\end{figure*}

To check whether the asymmetry in switching current is due to contact related issues, normal state $\frac{dV}{dI}$ is measured at different magnetic fields for up and down directional sweep of biasing current. Fig.~\ref{fig:figS9} shows the $\frac{dV}{dI}$ response at a particular $B$. It is clear from Fig.~\ref{fig:figS9} that for up and down direction biasing current sweep normal state differential resistance is not different within the noise limit. The asymmetry of switching current that we observe is not related to contacts.

\section{Thicknesses and perpendicular \textit{B} range for asymmetry tunability of the twisted devices}

We have performed AFM measurements to get the thicknesses of the individual BSCCO flakes. These measurements are done after we complete the electrical measurements on these devices. Thickness data are tabulated below. In the table, we also provide the areas of the twisted junctions for all the devices along with the perpendicular $B$ range over which we observe the switching current asymmetry tuning. The $B$ range for tuning asymmetry is always smaller for the $45\degree$ twisted devices than the $0\degree$ and $22\degree$ twisted devices.

\captionof{table}{Thickness, junction area, and \textit{B} range for asymmetry tunability of all the twisted devices.}

\begin{tabularx}{1\textwidth} { 
    | >{\raggedright\arraybackslash}X 
    | >{\raggedright\arraybackslash}X
    | >{\raggedright\arraybackslash}X
    | >{\raggedright\arraybackslash}X
    | >{\raggedright\arraybackslash}X
    | >{\raggedright\arraybackslash}X | }

    \hline
    Device No. & Twist angle & Top flake thickness & Bottom flake thickness & Junction area (\textmu m$^2$) & $B$ range for asymmetry tunability (\textmu T) \\
    \hline
    D1 & 45\degree & 24 nm & 13 nm & 340 & $\sim$ 20 \\ 
    \hline
    D2 & 0\degree & 50 nm & 30 nm & 290 & $\sim$ 300 \\ 
    \hline
    D3 & 45\degree & 60 nm & 30 nm & 154 & not measured \\ 
    \hline
    D4 & 0\degree & 32 nm & 24 nm & 367 & not measured \\ 
    \hline
    D5 & 22\degree & 65 nm & 25 nm & 450 & $\sim$ 300 \\ 
    \hline
    D6 & 45\degree & 45 nm & 34 nm & 1387 & $\sim$ 30 \\ 
    \hline
    D7 & 45\degree & 60 nm & 30 nm & 248 & $\sim$ 20 \\ 
    \hline
    D8 & 0\degree & 150 nm & 40 nm & 720 & $\sim$ 250 \\ 
    \hline
    D9 & 0\degree & 55 nm & 18 nm & 234 & $\sim$ 350 \\ 
    \hline
    D10 & 22\degree & 90 nm & 20 nm & 187 & $\sim$ 400 \\ 
    \hline
    
\end{tabularx}

\section{Measurement setup for field dependent asymmetry}

We have measured our devices in two different setups. Setup 1 is a closed-cycle cryogenic probe station (Lakeshore CRX-6.5 K) with a homemade electromagnet. Setup 2 is a wet cryostat (Oxford instrument) with an inbuilt superconducting magnet.

\begin{figure*}[h]
\centering

\includegraphics[width=15.24cm]{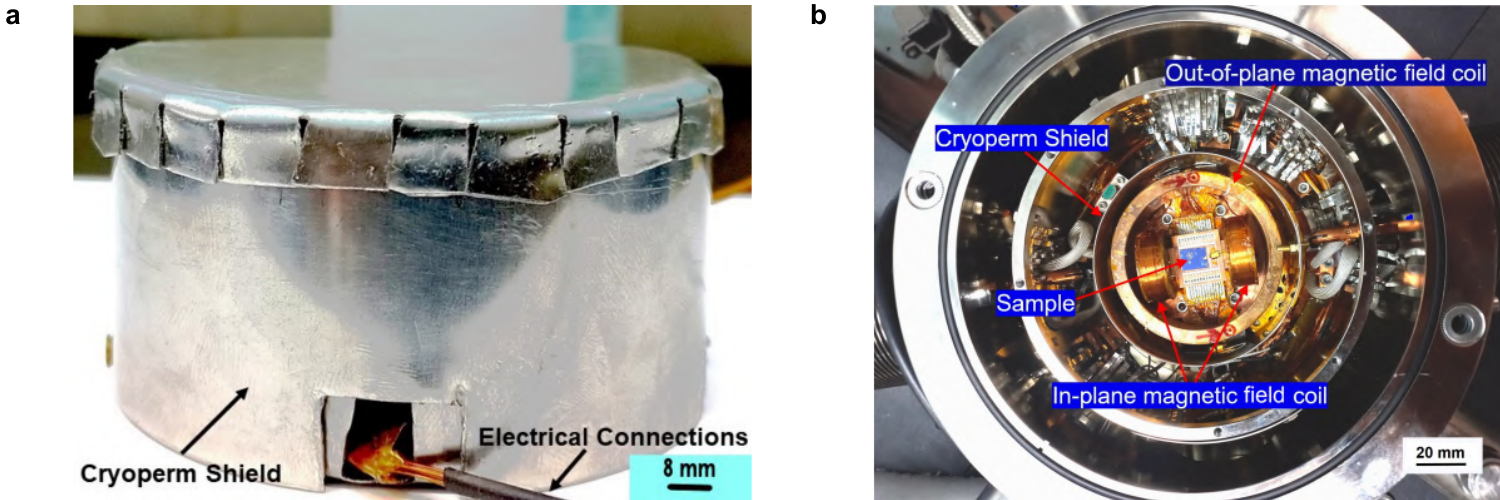}
\captionsetup{font=footnotesize}
\caption{ \label{fig:figS11}  \textbf{Magnetic coil setup and cryoperm shield in the Lakeshore probe station.} (a) The cryoperm shielding. It prevents stray magnetic fields from affecting the devices low temperatures. (b) Inside design of the Lakeshore probe station with magnetic coil and cryoperm shield. } 

\end{figure*}

The homemade electromagnetic coil in setup 1 is a solenoidal coil made up of 1200 turns of copper wire (diameter 0.2 mm) with an inner radius of 19.5 mm, an outer radius of 24.5 mm, and a width of 12 mm. The fabricated devices are mounted at the center of this homemade solenoid coil. The direction of the generated magnetic field is perpendicular to the plane of the BSCCO flake. We also installed coils for generating an in-plane magnetic field. The electromagnet along with the device is then covered from all sides using a cryoperm magnetic shield, which prevents any stray magnetic field from affecting the device. Fig.~\ref{fig:figS11}a shows the cryoperm magnetic shield. We have modified the inside design of a  closed-cycle cryogenic probe station (Lakeshore CRX-6.5 K) to accommodate the entire assembly of the coil and the shield, as shown in Fig.~\ref{fig:figS11}b. The 45\degree~twisted device presented in the main manuscript is measured in this homemade electromagnetic setup (setup 1) inside the probe station.

Setup 2 is the Oxford instrument wet cryostat where an inbuilt superconducting magnet generates the magnetic field perpendicular to the device plane. This setup has no magnetic shielding. The data of the 0\degree~and 22\degree~twisted devices, presented in the main manuscript, are taken in this setup 2.

To accurately determine the actual magnetic fields at the position of the device, we calibrate the magnetic fields in both setups. The calibration is done by the Bartington Mag-01H single-axis fluxgate magnetometer with Mag G probe DR2909 sensor. The sensor has a measuring range of 0 to 2~mT with a sensitivity of 1~nT. We place the sensor at the position of the device, perpendicular to the device plane. By sending a current through a Keithley source meter (model no. 2635A) we measured the field generated by the electromagnets. 


\section{Temperature dependence of $I_\text{s}^{+}$ and $I_\text{s}^{-}$}

\begin{figure*}[h]
\centering

\includegraphics[width=11.5cm]{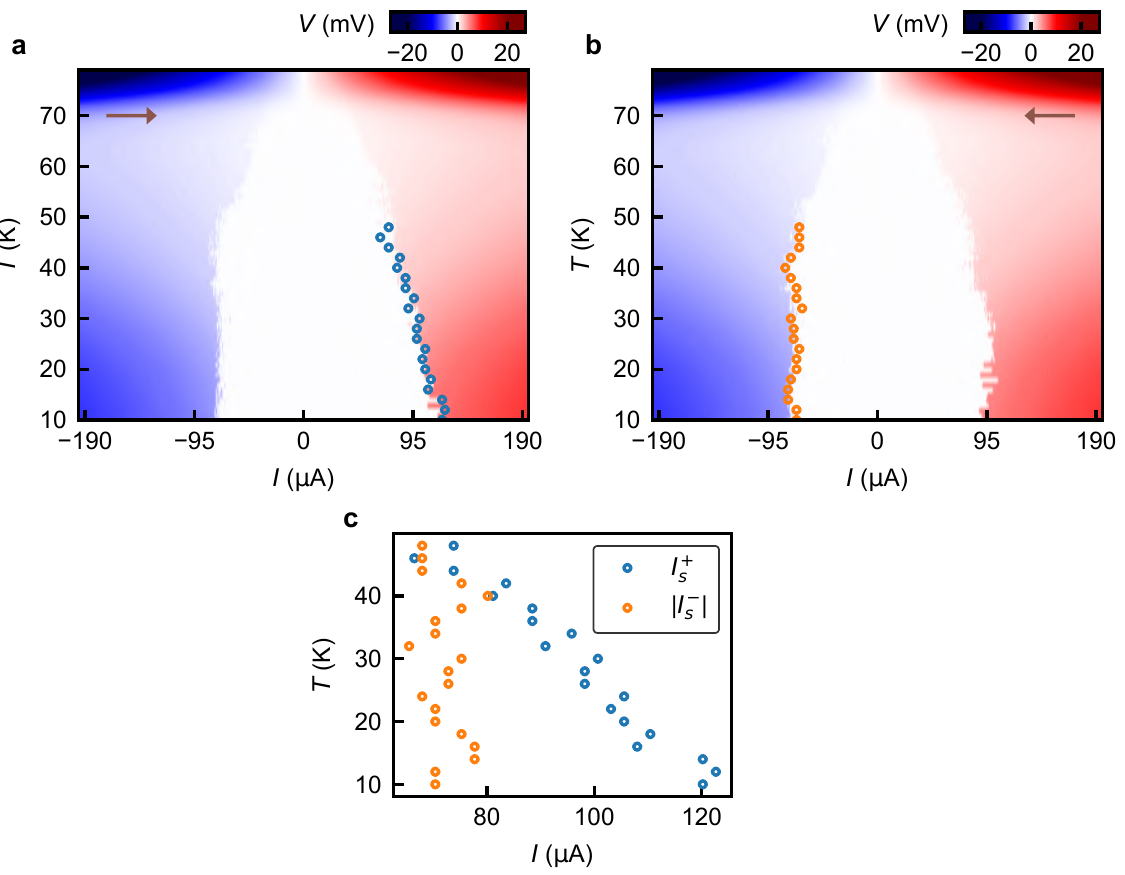}
\captionsetup{font=footnotesize}
\caption{ \label{fig:figS12}  \textbf{Temperature dependent positive and negative switching currents for $45\degree$ twisted device (D1).} (a), (b) Colorscale plot of $dc$ I-V with temperature for positive and negative sweep direction of bias current (indicated by the arrows), respectively. (c) Temperature-dependent $I_\text{s}^{+}$ and $I_\text{s}^{-}$ obtained from (a) and (b).} 

\end{figure*}

Fig.~\ref{fig:figS12}a, b show temperature dependent $dc$ I-V for both directions of bias current sweep for the $45\degree$ twisted device (D1). The switching currents at different temperatures for positive bias ($I_\text{s}^{+}$) are shown by open blue circles in Fig.~\ref{fig:figS12}a. The switching currents at different temperatures for negative bias ($I_\text{s}^{-}$) are shown by open orange circles in Fig.~\ref{fig:figS12}a. Fig.~\ref{fig:figS12}c shows temperature dependence of $I_\text{s}^{+}$ and $I_\text{s}^{-}$ simultaneously. This is data from the same device (D1) as Fig. 1 and Fig. 2 in the main manuscript. The data has been taken with zero current in the magnet coil (for producing a magnetic field in the setup). But as we mentioned in our manuscript (Caption in Fig. 2), there is a field offset of $\sim 5$ \textmu T in our measurement setup which accounts for the asymmetry between $I_\text{s}^{+}$ and $I_\text{s}^{-}$ even at zero externally applied $B$.

\section{Fraunhofer modulation of switching current -- a signature of Josephson tunneling}

\begin{figure*}[h]
\centering

\includegraphics[width=11.5cm]{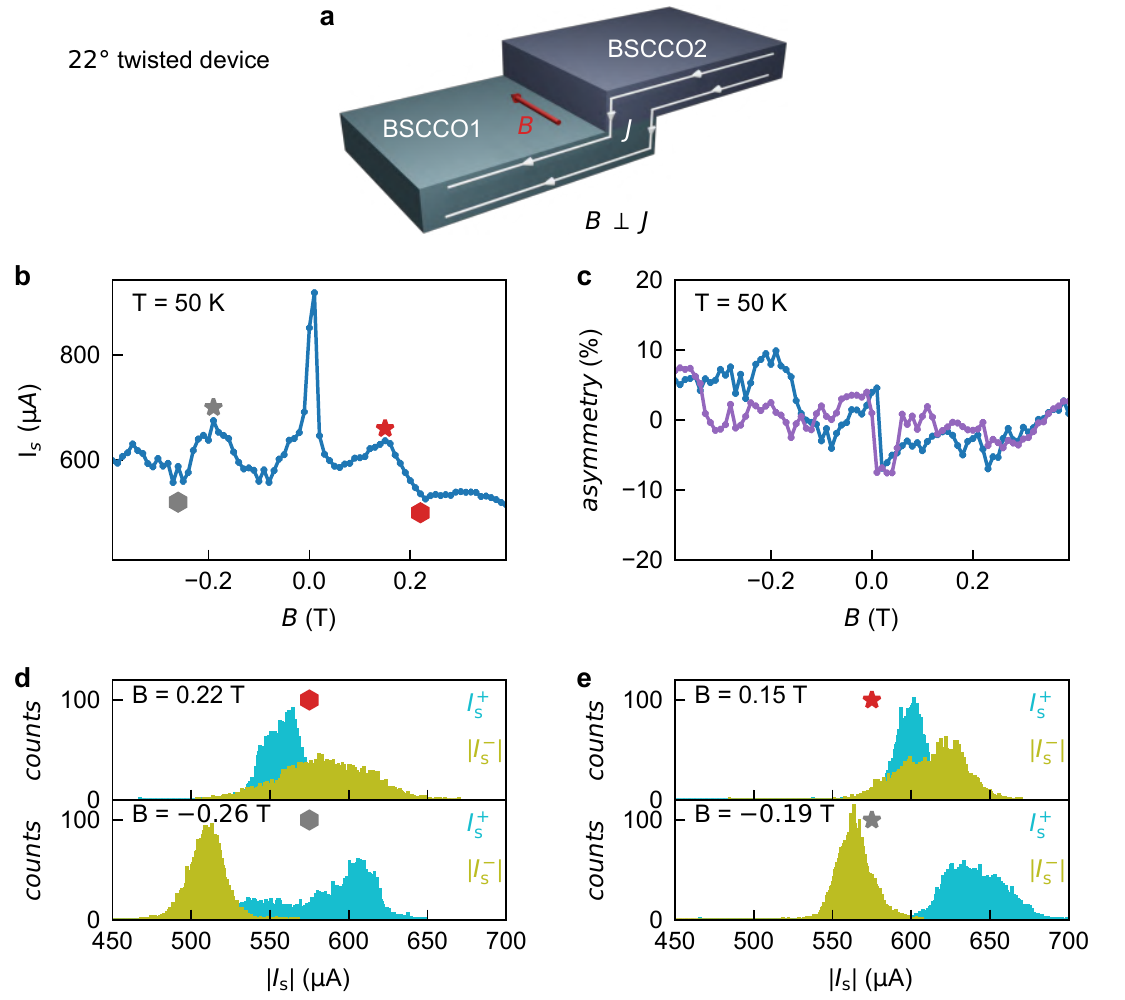}
\captionsetup{font=footnotesize}
\caption{ \label{fig:figS13}  \textbf{Modulation of switching current of the AJJ with an in-plane magnetic field ($J$ $\perp$ $B$).} (a) Schematic of Josephson current ($J$) and field ($B$) orientation at the junction. (b) Fraunhofer modulation of smallest switching current (attributed to AJJ) of a $22\degree$ twisted device (D5) with applied in-plane $B$. At each $B$, 100 switching events were recorded and averaged out to get the switching currents. (c) Calculated asymmetry factor from $I_\text{s}^{+}$ and $I_\text{s}^{-}$ as described in the main manuscript. (d), (e) Switching distributions of 10000 switching events for $I_\text{s}^{+}$ and $I_\text{s}^{-}$ at different $B$ marked by grey and red symbols in (b).} 

\end{figure*}

A strong signature of a Josephson junction is the Fraunhofer pattern-like modulation of its critical current in a magnetic field applied perpendicular to the Josephson current ($J \perp B$). For our $c$-axis twisted BSCCO devices, this will happen when the applied $B$ will be in-plane ($J \perp B$), as shown in Fig.~\ref{fig:figS13}a. We observe a Fraunhofer-like pattern of the smallest switching current (attributed to the artificial Josephson junction at the interface) with in-plane $B$ as shown in Fig.~\ref{fig:figS13}b at 50~K for a $22\degree$ twisted device (D5). This shows the Josephson coupling between the two twisted layers of BSCCO and the Josephson junction is in the short junction limit (Josephson penetration length $\lambda_{J} < W$; $W$ is the width of the Josephson junction).  We observe asymmetry between $I_\text{s}^{+}$ and $I_\text{s}^{-}$ in this configuration as well. However, we attribute this observed asymmetry to the contribution of the out-of-plane component of applied $B$ because of a small tilt between the sample plane and $B$ direction. It is very hard to perfectly align the sample plane with the in-plane $B$ and avoid any out-of-plane component of it. As we see in Fig.~\ref{fig:figS14}e and f (data from the same device as in Fig.~\ref{fig:figS13}) the asymmetry arises with very small out-of-plane B ($\sim$ 100 \textmu T). So even a misalignment of in-plane $B$ by a few degrees will have a significant out-of-plane $B$ component. Fig.~\ref{fig:figS13}d, e show distributions for 10000 switching events for $I_\text{s}^{+}$ and $I_\text{s}^{-}$ at different $B$ marked by red and grey star and hexagon in Fig.~\ref{fig:figS13}b. The shapes of the switching distributions for $I_\text{s}^{+}$ and $I_\text{s}^{-}$ are asymmetric.

\section{In-plane vs. out-of-plane magnetic field dependent switching current asymmetry}

\begin{figure*}[h]
\centering

\includegraphics[width=11.5cm]{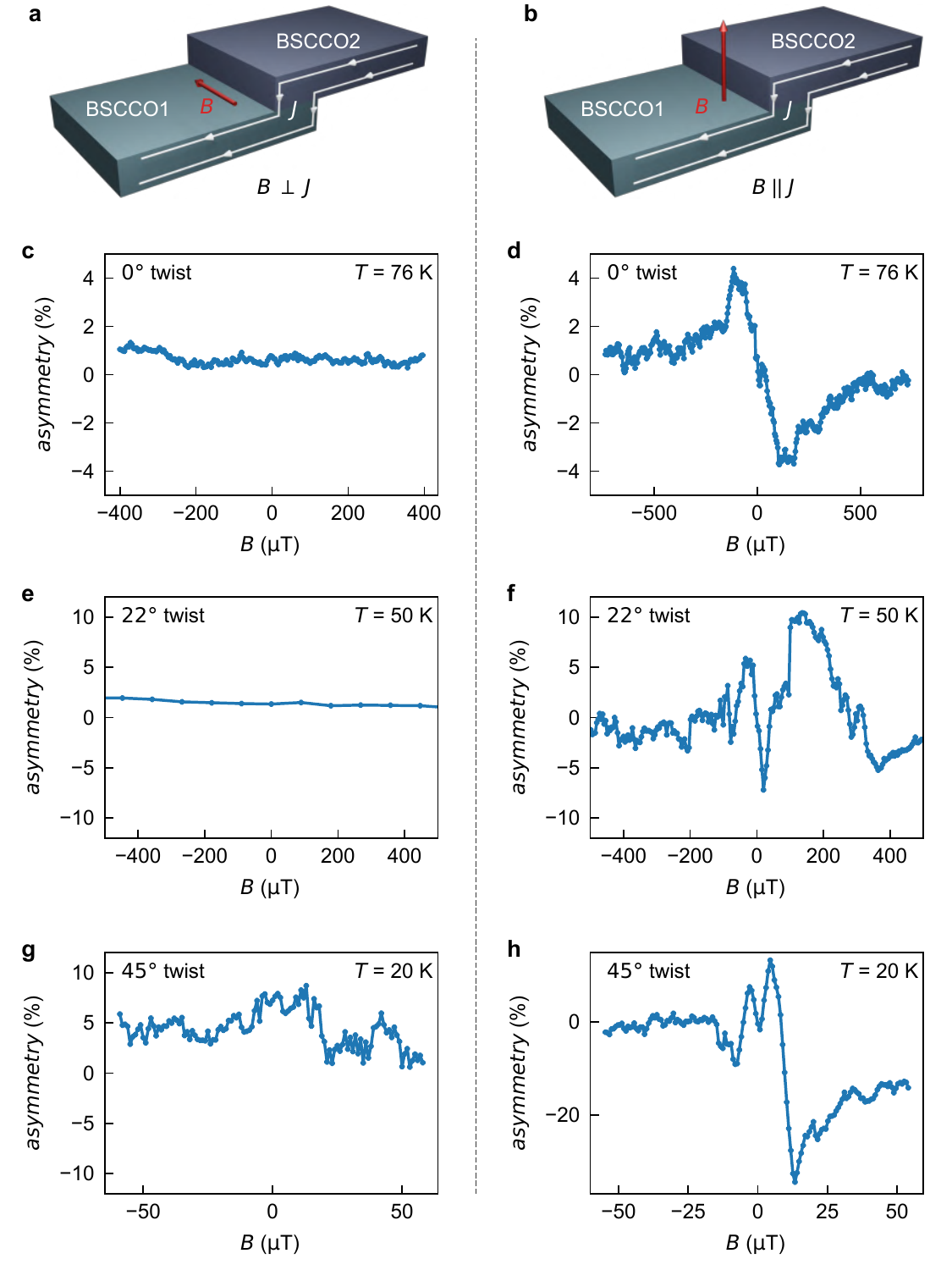}
\captionsetup{font=footnotesize}
\caption{ \label{fig:figS14}  \textbf{Switching current asymmetry for devices with different twist angles in presence of in-plane and out-of-plane B.} (a), (b) Shows two different configurations of Josephson current ($J$) and magnetic field ($B$). (c), (e), (g) asymmetry in $0\degree$ (D8), $22\degree$ (D5) and $45\degree$ (D6) twisted device with in-plane $B$. (d), (f), (h) asymmetry in $0\degree$ (D8), $22\degree$ (D5) and $45\degree$ (D6) twisted device with out-of-plane $B$.} 

\end{figure*}

Fig.~\ref{fig:figS14} summarizes the different behaviour of switching current asymmetry in devices with different twist angles in two different configurations of externally applied magnetic field ($B$). In out-of-plane $B$ (Fig.~\ref{fig:figS14}b), we observe large asymmetry in switching current for all twist angles (Fig.~\ref{fig:figS14}d, f, h). When we apply the same range of $B$ in-plane (Fig.~\ref{fig:figS14}a), we do not see a significant asymmetry in switching current (Fig.~\ref{fig:figS14}c, e, g). 

\section{Control experiment in pristine BSCCO without involving junction}

\begin{figure*}[h]
\centering

\includegraphics[width=12.7cm]{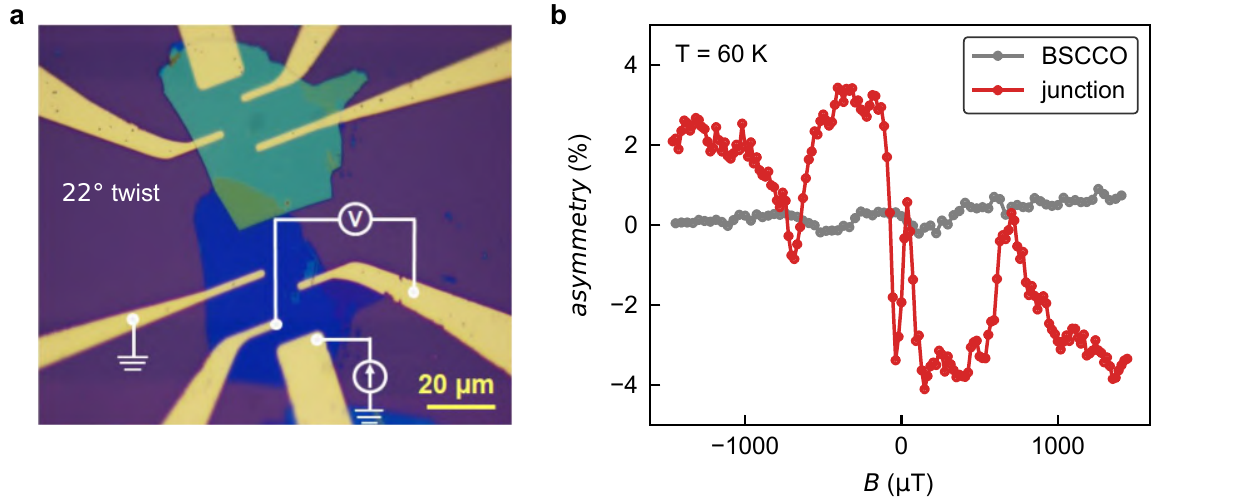}
\captionsetup{font=footnotesize}
\caption{ \label{fig:figS15}  \textbf{Control experiment in pristine BSCCO without involving junction.} (a) Optical micrograph of a $22\degree$ twisted device (D10) with measurement protocol to probe pristine BSCCO without involving junction. (b) Comparison plot of asymmetry between the junction and pristine BSCCO at 60~K. The data is consistent with Extended Data Fig. 2.} 
\end{figure*}

In Fig.~\ref{fig:figS15}, we present a control experiment in extended pristine BSCCO away from the twisted junction. This control experiment is similar to the control experiment provided in Extended Data Fig. 2 except that the bias current now flows only in the pristine BSCCO and not through the junction. Fig.~\ref{fig:figS15}a shows the measurement protocol for switching current asymmetry in the pristine BSCCO. Fig.~\ref{fig:figS15}b shows simultaneously the asymmetry arising from the junction and pristine BSCCO. Similar to the data in Extended Data Fig. 2, we observe very small asymmetry ($< 1\%$) in pristine BSCCO. This control experiment along with the one provided in Extended Data Fig. 2 show that there is a negligible asymmetry in the pristine BSCCO flake in presence of an out-of-plane B.   

\section{Theoretical scenario}

\subsection{Origin of diode behavior}

We provide additional background on the theoretical scenario sketched in the main text. 
Due to its layered structure, BSCCO has strongly asymmetric magnetic response, with $\lambda_{ab} \sim 0.2\mu$m much smaller than the width of the sample  
and $\lambda_c\sim 100\mu$m much larger than the height of the flakes. The magnetic field applied along the $c$-axis will thus enter in stacks of pancake vortices. The AJJ realizes an extended Josephson contact (in the $xy$-plane) with Josephson current 
\begin{equation}
    I = \int dx dy J_c \sin\varphi(x,y),
    \label{eq:Fraunhofer}
\end{equation}
where $\varphi(x,y)$ denotes the gauge-invariant phase difference across the junction as a function of position. Equation (\ref{eq:Fraunhofer}) assumes a spatially independent coupling strength (and hence $J_c$)  across the contact as well as weak coupling so that the Josephson current is sinusoidal in the phase difference.   

In general, the phase difference depends on the externally applied field as well as self fields generated by the Josephson current. We estimate that self-fields are negligible since the junction width is smaller than the Josephson penetration depth $\lambda_J=\sqrt{\phi_0/4\pi\mu_0 j_c d}\sim 100\mu$m (with sample thickness $d$, flux quantum $\phi_0=h/2e$, and critical current density $j_c$ taken at a twist angle of $45^\circ$). This is consistent with the observed Fraunhofer-like interference at large in-plane fields (Fig.~\ref{fig:figS13}). Thus, the phase difference is controlled by the applied magnetic field and remains uniform across the junction as long as the field is pointing along the $c$-axis only. This is true even if the flux enters in $c$-axis-aligned vortex stacks, for which the phase configurations of top and bottom flakes would be identical.

\begin{figure*}
\centering

\includegraphics[width=8.9cm]{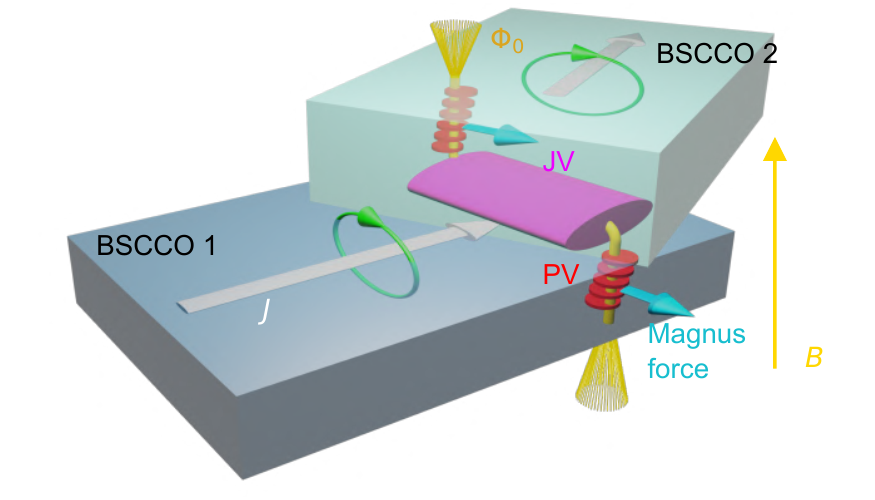}
\caption{ \label{fig:fig17}  \textbf{Mechanism of diode effect.} Schematic of a twisted BSCCO junction with possible vortex configuration. Out-of-plane $B$ induces stacks of pancake vortices (PV) which can be pinned at different locations in top and bottom flake and connect through a Josephson vortex (JV). The difference in vortex configurations along with the in-plane field of JV result in a spatially nonuniform gauge-invariant phase difference across the AJJ and thus Fraunhofer-like interference. Changes in the vortex configurations due to the Magnus force exerted by the bias current ($J$) will differ depending on bias direction, resulting in diode behavior. We estimate that self-fields of the bias current (green loops) contribute only weakly to the asymmetry.
} 
\end{figure*}

The phase difference is no longer spatially uniform when the vortex stacks are not passing vertically through the entire sample. 
At the AJJ,  
the Josephson coupling and hence the coupling between pancake vortices is weakest. Moreover, top and bottom flakes have uncorrelated defect contributions, favoring pinning of vortex stacks at different sample locations. As a result, magnetic flux must exit at the AJJ within the $ab$-plane or connect between vortex stacks of the top and bottom layers \cite{golubov_interaction_1992,blatter_vortices_1994}. (Effectively, the latter situation would correspond to the formation of a Josephson vortex at the AJJ, although it should be kept in mind that the flux is only weakly concentrated at the interface as the height of the flakes is small compared to $\lambda_c$.) This results in a spatially dependent phase difference and thus Fraunhofer-like interference [see Eq.\ (\ref{eq:Fraunhofer})], both due to the incongruent phase configurations associated with the vortices in top and bottom flakes and due to the  magnetic field along the $ab$-plane. 

Josephson diode behavior appears due to self-field effects or current-induced changes in the vortex configurations 
(Magnus force). The sample geometry is such that the bias current is effectively flowing within the $ab$-plane. Thus, self fields contribute to the flux penetrating the junction parallel to the $ab$-plane. Moreover, the bias current exerts Magnus forces on the vortex stacks in the upper and lower flake. Importantly, both the self field and the Magnus force change sign with the direction of current bias and thus affect the Fraunhofer-like interference asymmetrically. The changes in the interference pattern can be substantial when the vortex configuration changes. In contrast, we estimate below (see Sec.\ \ref{Sec:sfe}) that self-field effects tend to be smaller than the observed asymmetries, a fact which is consistent with the large Josephson length. We thus conclude that the diode effect is dominated by  vortex rearrangements in the presence of Magnus forces exerted by the bias current. 

As described in the main text, this scenario is consistent with the principal experimental observations. In the next section, we give more details on the switching-current distributions. 

\subsection{Switching current distribution}
\label{Sec:distribution}

The switching dynamics of the Josephson diode can be described within the model of a resistively and capacitively shunted Josephson junction (RCSJ model), see, e.g., \cite{tinkham_introduction_2004,steiner_diode_2022}. Within this model, the phase dynamics in the presence of a bias current $I_b$ is described by the Langevin equation
\begin{equation}
  \frac{\hbar C}{2e}\ddot \gamma + \frac{\hbar}{2eR}\dot \gamma + I(\gamma) + \delta I = I_b 
\end{equation}
for the phase difference $\gamma$. Here, $C$ is the junction capacitance, $R$ accounts for quasiparticle dissipation, $\delta I$ is the associated Johnson-Nyquist noise with correlator $\langle \delta I(t)\delta I(t')\rangle = (2k_BT/R)\delta(t-t')$, and $I(\gamma)$ denotes the current-phase relation.  
The Johnson-Nyquist noise 
induces a distribution of switching currents $I_s$. The  average is  reduced relative to the critical current (maximum of current-phase relation) and the variance of the distribution is controlled by  (the effective) temperature, $\Delta I_s \simeq 2e k_BT/\pi\hbar$ (see Sec.\ \ref{Sec:scd} below).   Equation (\ref{eq:Fraunhofer}) gives a current-phase relation 
\begin{equation}
    I(\gamma) = I_0(B,I_b)\sin\gamma,
\end{equation}
where $\gamma$ is the average phase difference across the junction (i.e., a constant offset of $\varphi(x,y)$).
According to the discussion in the previous section, the critical current depends on the applied magnetic field $B$ as well as the bias current $I_b$. 

We note that for $T\simeq 50$K, one estimates $\Delta I_s \sim 1\mu$A. This is close to, but smaller than the experimentally observed values. 
Additional nonthermal noise contributions to the Langevin current $\delta I$ increase the width, but cannot explain the observed multiple humps and  asymmetries in the width of the switching current distribution. Both of these can be understood as a consequence of multistabilities of the vortex configurations at fixed $B$ and $I_b$. Multihump distributions reflect multistability of vortex configuations with sufficiently  different critical currents. Asymmetries  reflect the distinct (multistabilities of the) vortex configurations for the two bias corrections, which also underlie the observed asymmetry of the average switching current. 

Asymmetric widths of the switching current distributions can also appear as a consequence of non-sinusoidal current-phase relations, see Sec.\ \ref{Sec:scd}. An asymmetry arises for washboard potentials in which the minimum is  located asymmetrically between neighboring maxima. A similar effect can occur as a result of self fields, which effectively act as a current-induced phase offset of the current-phase relationship. While this can in principle contribute to asymmetric widths, this mechanism  cannot explain the multi-hump distributions and is presumably weak along with higher harmonics of the current-phase relationship and/or the self-field effects.

\subsection{Self-field effects due to the bias current}
\label{Sec:sfe}

We estimate the asymmetry due to self-field effects of the bias current. The sample geometry is such that the bias current effectively flows within the $ab$-plane. A thin-plate conductor in the $xy$-plane (thickness $d$) with a uniform current density $j$ flowing in the $x$-direction produces a magnetic field
\begin{equation}
   B_y = \pm \frac{1}{2}\mu_0 j d 
\end{equation}
in the $y$-direction, which is uniform above and below the plate, with opposite signs. 
Viewing the entire sample as such a thin plate conductor (on scales large compared to its thickness), we then find that it induces a field $B_y \simeq \frac{\mu_0 I_\mathrm{bias}}{2L_y}$ sufficiently far from the sample plane, where $L_y$ is the sample width in the $y$-direction and $I_\mathrm{bias}$ is the current bias. 

We are interested in the induced field in the junction region. For the purpose of an estimate, we simplify considerations by assuming that the system and the magnetic-field configuration is uniform in the $y$-direction.
Then, the Josephson current in the $z$-direction can be expressed as 
\begin{equation}
    j_z(x) = J_c \sin(\gamma + \frac{2\pi\phi(x)}{\phi_0} ),
\end{equation}
where we assume that the junction extends between $-L_x/2$ and $L_x/2$ in the $x$-direction (i.e., $\gamma$ is the phase difference at the center of the junction). Here, $\phi(x)$ denotes the flux penetrating the junction in the $y$-direction between $x=0$ and $x$. Due to the inhomogeneous field distribution originating from the vortex configuration, this will usually not be a linear function of $x$. 

We want to compute the currents in the upper and lower electrodes. We first note that the currents are approximately uniform in the $z$-direction, since the thickness of the flakes is small compared to the London penentration depth. The cross section at any $x$ must be passed by the current $I_\mathrm{bias}$. Correspondingly, we write the current in the $x$-direction in the lower flake as $I(x)$ and in the upper flake as $I_\mathrm{bias}-I(x)$. Continuity implies
\begin{equation}
    dI(x) = j_z(x) L_y dx.
\end{equation}
This should be supplemented by the boundary conditions $I(-L_x/2) = 0$ and $I(L_x/2)=I_\mathrm{bias}$. 

We can now obtain 
\begin{equation}
    I(x) = L_y \int_{-L_x/2}^x dx' j_z(x')
\end{equation}
with the condition that 
\begin{equation}
    I_\mathrm{bias} = L_y \int_{-L_x/2}^{L_x/2} dx' j_z(x').
\end{equation}
Note that this condition basically determines the phase difference across the junction. We enclose, say, the lower flake by an Amperian loop to obtain the induced magnetic field within the junction. This gives 
\begin{equation}
    B_\mathrm{jct}(x) = 
    - \frac{\mu_0 I_\mathrm{bias}}{2L_y} + \frac{\mu_0 I(x)}{L_y}.
\end{equation}
This field in turn affects the phase difference, so that we should use
\begin{equation}
    \phi(x) = d_\mathrm{eff} \int_{0}^{x} dx' [B_\mathrm{ext}(x')+B_\mathrm{jct}(x')]. 
\end{equation}
Here, the effective thickness $d_\mathrm{eff}$ is of the order of the thickness of the sample in the $z$-direction. 

Estimating the contribution $\delta\phi$ of the induced field to the flux penetrating the junction (total sample thickness of $50$nm, a width of 20$\mu$m, and bias current $I_\mathrm{bias}\sim 50\mu$A), we estimate $\delta\phi \lesssim 10^{-2}\phi_0$, which would correspond to an asymmetry of the order of a percent, considerably smaller than observed in the experiment. 

\subsection{Width of switching-current distribution}
\label{Sec:scd}

We derive the switching-current distribution accounting for general current-phase relationships. Our calculation extends the calculation of the average switching current presented in Ref. \cite{steiner_diode_2022} to the distribution function. Our notation, including the definition of dimensionless units, follows this reference. 

The (dimensionless)
switching rate is given by \cite{Haenggi1990,steiner_diode_2022}
\begin{equation}
    \gamma_\mathrm{sw}
    = \frac{\varepsilon_d \omega_0}{2\pi \theta}
    \exp\left\{-\frac{\varepsilon_B}{\theta}\right\},
    \label{eq:gammatr}
\end{equation}
where $\varepsilon_B$ is the barrier height, $\epsilon_d$ is the energy dissipated along the separatrix motion, and $\omega_0$ is the plasma frequency. All quantities, including temperature $\theta$, are in dimensionless units, see \cite{steiner_diode_2022}. We consider a linear ramp of the bias current $i_b$ from zero (ramp rate $a>0$),
\begin{equation}
    i_b(\tau)=\pm a\tau.
    \label{eq:itau}
\end{equation}
The sign indicates the direction of the ramp. We denote by $P(\tau)$ the probability that  the system is still in the trapped state at (dimensionless) time $\tau$. This probability
obeys the rate equation
\begin{equation}
    \frac{dP}{d\tau} = -\gamma_\mathrm{sw}(\tau)P.
\end{equation}
The escape rate $\gamma_\mathrm{sw}(\tau)$ out of the trapped state is time dependent, since the barrier height is a function of the bias current $i_b$. Solving for $P(\tau)$ with initial condition $P(\tau=0)=1$ gives
\begin{equation}
    P(\tau) = \exp\left\{-\int_0^\tau d\tau'\, 
    \gamma_\mathrm{sw}(\tau')
    \right\}.
\end{equation}
We use Eq.\ (\ref{eq:itau}) to replace time by current,
\begin{equation}
    P(i_b) =  \exp\left\{-\frac{1}{a}\int_0^{\abs{i_b}} di\, \gamma_\mathrm{sw}(\pm i)
    \right\}.
\end{equation}
The switching rate $\gamma_\mathrm{sw}$ depends exponentially on the barrier height $\varepsilon_B=\varepsilon_B(i_b)$, which decreases with bias current. At low temperatures, the integral in the exponent is thus dominated by the upper limit. This gives
\begin{equation}
    P(i_b) \propto  \exp\left\{-\frac{\theta \gamma_\mathrm{sw}(i_b)}{a}\left|\frac{di_b}{d\varepsilon_B} \right|
    \right\}
    \label{eq:PIB}
\end{equation}

The switching probability $p_\mathrm{sw}(i_b)$ can be extracted from $P$ as 
\begin{equation}
   p_\mathrm{sw}(i_b) = \mp  \frac{dP(i_b)}{di_b} , 
\end{equation}
which yields
\begin{equation}
   p_\mathrm{sw}(i_b) = \frac{\gamma_\mathrm{sw}(i_b)}{a} 
   P(i_b).
\end{equation}
We expect $p_\mathrm{sw}(i_b)$ to be a sharply peaked function, which we approximate as a Gaussian. 

This is implemented by writing
\begin{equation}
p_\mathrm{sw}(i_b) \propto \frac{1}{a} 
       \exp\left\{-f(i_b)\right \}
\end{equation}
with
\begin{equation}
    f(i_b)
    = \frac{\theta \gamma_\mathrm{sw}(i_b)}{a}\left|\frac{di_b}{d\varepsilon_B} \right| - \ln \gamma_\mathrm{sw}(i_b).
\end{equation}
We expand $f(i_b)$ about its minimum at $i_\mathrm{sw}$, 
\begin{equation}
    f(i_b) \simeq f(i_\mathrm{sw})
    + \frac{1}{2}f''(i_\mathrm{sw})(i_b-i_\mathrm{sw})^2 .
\end{equation}
We have
\begin{equation}
    f'(i_b) = \left\{ \frac{\theta}{a} \left|\frac{di_b}{d\varepsilon_B} \right| -\frac{1}{\gamma_\mathrm{sw}(i_b)}
    \right\}
    \frac{d\gamma_\mathrm{sw}}{di_b}
\end{equation}
Thus, the condition $f'(i_\mathrm{sw})=0$ yields
\begin{equation}
    \gamma_\mathrm{sw}(i_\mathrm{sw}) = \frac{a}{\theta} \left|\frac{di_b}{d\varepsilon_B} \right|^{-1} 
    \label{eq:gswisw}
\end{equation}
and 
\begin{equation}
   f(i_\mathrm{sw}) = 1 - \ln \left(\frac{a}{\theta} \left|\frac{di_b}{d\varepsilon_B} \right|^{-1}\right). 
\end{equation}
Furthermore, we have
\begin{eqnarray}
    f''(i_\mathrm{sw})
    &\simeq& \frac{1}{\gamma^2_\mathrm{sw}(i_\mathrm{sw})} 
    \left(
    \left.\frac{d\gamma_\mathrm{sw}}{di_b}\right|_{i_b=i_\mathrm{sw}}\right)^2 
    \nonumber\\
    &\simeq& 
    \frac{1}{\theta^2}\left(\left.\frac{d\varepsilon_B}{di_b}\right|_{i_b=i_\mathrm{sw}}\right)^2.
\end{eqnarray}
We denote the phase differences at which the tilted washboard potential becomes maximal (minimal) as $\varphi_\mathrm{max}$ ($\varphi_\mathrm{min}$).
Using
\begin{eqnarray}
    \epsilon_B &=& u(\varphi_\mathrm{max}) - u(\varphi_\mathrm{min})
    \nonumber\\
    &=& u_0(\varphi_\mathrm{max}) - u_0(\varphi_\mathrm{min})
    -i_b(\varphi_\mathrm{max}-\varphi_\mathrm{min}),
\end{eqnarray}
we have 
quite generally
\begin{equation}
    \left|\frac{d\varepsilon_B}{di_b}\right| = |\varphi_\mathrm{max}-\varphi_\mathrm{min}|.
\end{equation}
We then find
\begin{equation}
 p_\mathrm{sw}(i_b) = \frac{\left|\varphi_\mathrm{max}^\pm-\varphi_\mathrm{min}\right|}{\sqrt{2\pi}\theta} 
   \exp\left\{-\frac{(\varphi_\mathrm{max}^\pm-\varphi_\mathrm{min})^2}{2\theta^2}(i_b-i_\mathrm{sw}^\pm)^2\right \} . 
\end{equation}
Here, we have fixed the numerical prefactor by normalization. We have added a superscript to $\pm$ to $\varphi_\mathrm{max}$, since the relevant maximum depends on current direction. Similarly, we indicated the dependence of the average switching current $i_\mathrm{sw}^\pm$ on the direction of the bias current. Finally reverting to dimensional variables (written in capitals), we obtain
\begin{equation}
 p_\mathrm{sw}(I_b) = \frac{\left|\varphi_\mathrm{max}^\pm-\varphi_\mathrm{min}\right|}{\sqrt{2\pi}(2ek_BT/\hbar)} 
   \exp\left\{-\frac{(\varphi_\mathrm{max}^\pm-\varphi_\mathrm{min})^2}{2(2e k_B T/\hbar)^2}(I_b-I_\mathrm{sw}^\pm)^2\right \} . 
\label{eq:DistSwitchCur}
\end{equation}
As long as the Langevin fluctuations reduce the average switching current significantly below the critical current, the tilt of the washboard potential is small and we can approximate $\left|\varphi_\mathrm{max}^\pm-\varphi_\mathrm{min}\right|\sim \pi$. This gives the estimate for the width $\Delta I_s$ of the switching current quoted above. For general tilts, this provides a lower limit to the width of the switching current distribution, which may moreover become asymmetric (but not multi-humped) for asymmetric $I_\mathrm{sw}^\pm$. 


\end{document}